\documentclass{article}

\usepackage{arxiv}

\usepackage[utf8]{inputenc} % allow utf-8 input
\usepackage[T1]{fontenc}    % use 8-bit T1 fonts
\usepackage{hyperref}       % hyperlinks
\usepackage{url}            % simple URL typesetting
\usepackage{booktabs}       % professional-quality tables
\usepackage{amsfonts}       % blackboard math symbols
\usepackage{nicefrac}       % compact symbols for 1/2, etc.
\usepackage{microtype}      % microtypography
\usepackage{lipsum}
\usepackage{graphicx}
\usepackage{amsmath}
\usepackage{breqn}
\graphicspath{ {./images/} }
\usepackage{natbib}
\usepackage{algorithm}
\usepackage{tabularx}
\usepackage{algpseudocode}

\title{TENSOR TRAIN DECOMPOSITION-BASED 3D IMPLICIT FULL WAVEFORM INVERSION WITH MULTI-SCALE STRUCTURAL SIMILARITY}

\author{
 Liangsheng He \\
  State Key Laboratory of Deep Earth Exploration and Imaging \\
  College of GeoExploration Science and Technology, Jilin University\\
  Changchun, Jilin, 130021, China \\
   \And
Chao Song \\
  State Key Laboratory of Deep Earth Exploration and Imaging \\
  College of GeoExploration Science and Technology, Jilin University\\
  Changchun, Jilin, 130021, China \\
  \And
Tiansheng Chen \\
  Petroleum Exploration and Production Research Institute, Sinopec\\
  Beijing, 100083, China \\
  \And  
 Tao Liu \\
  Petroleum Exploration and Production Research Institute, Sinopec\\
  Beijing, 100083, China \\
  \And  
 Cai Liu \\
  State Key Laboratory of Deep Earth Exploration and Imaging \\
  College of GeoExploration Science and Technology, Jilin University\\
  Changchun, Jilin, 130021, China \\
}

\begin{document}
\maketitle

\begin{abstract}
Three-dimensional full waveform inversion (3DFWI) is a powerful technique for reconstructing high-resolution subsurface velocity models. However, its application is often limited by high memory requirements, computational costs, and sensitivity to cycle skipping. To overcome these challenges, we propose a novel tensor train (TT) decomposition-based 3D implicit full waveform inversion framework (TT-3DIFWI) combined with a multi-scale structural similarity (M-SSIM) objective function. In this framework, the 3D velocity model is represented by TT decomposition as a product of a series of low-rank core tensors. Then, three axis-specific implicit neural network representations (INR) based on one-dimensional vector coordinates as input are constructed to predict these core tensors, rather than directly predicting the velocity model. This INR reparameterization method based on TT decomposition can significantly reduce the memory consumption of INR training while maintaining the accuracy and resolution of the 3D velocity model reconstruction. Meanwhile, the low-rank structure of TT decomposition also ensures the structural consistency of the reconstruction velocity, thereby improving the accuracy and continuity of the inversion result. Furthermore, the M-SSIM objective function can compare the multi-scale structural differences between predicted and observed data, and utilize the ultra-low frequency features to reduce cycle skipping. Numerical experiments on synthetic and challenging land datasets demonstrate that TT-3DIFWI with M-SSIM achieves accurate and continuous velocity reconstruction, even with poor initial models or missing low-frequency data. 
\end{abstract}


\section{Introduction}
Full waveform inversion (FWI) is a powerful seismic imaging technique that reconstructs high-resolution subsurface velocity models by minimizing the misfit between observed and predicted seismic data \cite{tarantolaInversionSeismicReflection1984}. Owing to its ability to utilize the full wavefield information, FWI has been widely recognized as one of the most promising approaches for high-resolution subsurface characterization \cite{virieuxOverviewFullwaveformInversion2009, brenders2007full, shen2018full}. However, FWI is inherently a highly nonlinear optimization problem. When the traveltime shift between predicted and observed waveforms exceeds half a cycle, the inversion process may converge toward an incorrect local minimum, a phenomenon commonly referred to as cycle skipping \cite{virieuxOverviewFullwaveformInversion2009}. This issue becomes particularly severe when the initial velocity model is far from the true model or when low-frequency components are missing from the seismic data \cite{tarantolaInversionSeismicReflection1984, wang2009reflection, warner2013anisotropic, chi2015correlation, wang2019full, song2021wavefield}.

To mitigate cycle skipping and improve the stability of inversion, various strategies have been proposed. These include traveltime-based inversion \cite{luo1991wave, wangFrequencydomainWaveequationTraveltime2021}, migration-based velocity analysis \cite{sava2004wave, symes2008migration, alaliEffectivenessPseudoinverseExtended2020}, reflection-based waveform inversion \cite{xu2012full, wu2015simultaneous, yaoReviewReflectionwaveformInversion2020}, and envelope-based inversion {chi2014full, wuSeismicEnvelopeInversion2014}. Such approaches aim to construct more reliable initial velocity models or enhance low-frequency information in seismic data, thereby reducing the likelihood of convergence to local minima.

In recent years, deep learning (DL) technology has been widely applied in FWI to improve inversion efficiency and robustness against cycle skipping \cite{karniadakis2021physics, wu2023sensing, schuster2024review, song2026physics}. Early DL-based FWI primarily focused on data-driven frameworks, where neural networks (NNs) are trained to learn a direct mapping from seismic data to subsurface velocity models. Researchers adopted various DL architectures, including encoder–decoder based convolutional neural networks (CNNs) \cite{wang2018velocity,yang2019deep,wu2019inversionnet,feng2021multiscale}, recurrent neural networks (RNNs) \cite{richardson2018seismic,adler2019deep,fabien2020seismic}, and generative adversarial networks (GANs) \cite{zhang2020data,mosser2020stochastic,wang2022seismic}, to extract useful features from seismic records and predict velocity structures. In these approaches, the inversion problem is reformulated as a supervised learning task that relies on large datasets of paired seismic data and velocity models. These data-driven methods can significantly accelerate the inversion process after training and can capture complex patterns in seismic data through the strong representation capability of NNs. However, their performance heavily depends on the quality and diversity of the training data. When applied to unseen geological structures, different acquisition geometries, or field data, their prediction accuracy may decrease noticeably \cite{yang2019deep, wu2019inversionnet}. In contrast, physics-driven DL-FWI incorporates the wave equation into the NNs training process, thereby eliminating the dependence on large-scale labeled datasets \cite{song2021wavefield, sun2023implicit, zhang2023multilayer, yang2023fwigan}. Among various physics-driven methods, implicit full waveform inversion (IFWI) stands out by using implicit neural representation (INR) to reparameterize the velocity model, thereby improving the accuracy and continuity of the inversion results\cite{sun2023implicit, yang2025gabor}. In IFWI, NNs are employed to represent spatially continuous fields, modeling the velocity as a function of spatial coordinates. Compared with conventional grid-based parameterizations, INR provides a more compact representation and introduces implicit regularization, which improves the smoothness of the reconstructed model.

With the increasing demand for accurate three-dimensional (3D) subsurface characterization \cite{ben2008velocity,raknes2015three,amestoy2016fast,wang2021anisotropic,zhang20233,he2026m}, three-dimensional IFWI (3DIFWI) has become an important research direction. However, the transition of IFWI from two-dimensional (2D) to large-scale 3D scenarios introduces significant computational bottlenecks. In a 3D environment, the NNs must ingest the spatial coordinates $(x, y, z)$ for every grid point in the volume. During the forward and backward propagation stages of NNs, a large number of intermediate activation values are generated. For high-resolution 3D models, the memory requirement for storing these intermediate variables grows exponentially with the model dimensions, often exceeding the capacity of standard graphics processing unit (GPU) hardware. Furthermore, IFWI primarily relies on implicit regularization and spectral bias inherent in neural network architectures to improve inversion results. This single constraint is usually insufficient to guarantee the stability of the inversion process, resulting in poor consistency of subsurface structure in the IFWI inversion result. Furthermore, the choice of the objective function is crucial to the success of IFWI. Traditional IFWI typically uses the $l_2$ norm to construct the objective function, which measures the point-by-point amplitude difference between observed and predicted data. However, this point-by-point comparison is highly sensitive to phase shifts and can still easily lead to period skipping when the initial model is inaccurate or lacks low-frequency information.

To address these challenges, this study proposes a novel tensor train (TT) decomposition 3DIFWI framework (TT-3DIFWI) combined with a multi-scale structural similarity (M-SSIM) objective function. In this framework, the 3D velocity model is represented using TT decomposition as a sequence of low-rank core tensors. These tensors are predicted through the proposed axis-specific implicit neural representations (INRs), which require only coordinate vectors along individual spatial directions rather than the full spatial coordinates of the velocity model, thereby significantly reducing memory consumption. Meanwhile, the inherent low-rank structure of TT decomposition promotes structural consistency in the reconstructed velocity model, improving both the accuracy and continuity of the inversion results. Furthermore, the M-SSIM objective function measures the multi-scale structural differences between the observed and predicted seismic data, enhancing the sensitivity of inversion to structural features and improving robustness against cycle skipping. Experiments on synthetic and challenging land datasets demonstrate that the proposed TT-3DIFWI with M-SSIM can accurately reconstruct velocity models with improved structural continuity, even when the initial model is inaccurate or low-frequency components are missing. These results highlight the feasibility, effectiveness, and robustness of the proposed framework for large-scale 3D seismic inversion under limited computational resources.

\section{Methodology}

This paper proposes a novel TT decomposition-based 3D-IFWI (TT-3DIFWI) framework with an M-SSIM objective to achieve memory-efficient and structurally consistent speed inversion. In this section, we first briefly review the formulation of conventional 3DFWI. Then, we introduce the 3DIFWI framework, where the subsurface velocity model is represented using coordinate-based NNs. Next, we introduce the TT-3DIFWI method. This method decomposes the 3D velocity model into a series of low-rank core tensor components. We predict these core tensors using axis-specific INRs and reconstruct the velocity field using TT representations. Finally, we introduce the M-SSIM objective function and discuss its advantages in improving structural consistency and reducing cycle skipping effects during the inversion process.

\subsection{3DFWI}\hfill

The 3D acoustic wave equation is expressed as follows:
\begin{equation}
\label{deqn_ex1a}
\dfrac{1}{v^2}\dfrac{\partial^2 u(\mathbf{x},t)}{\partial t^2} - \nabla^2 u(\mathbf{x},t)= s(\mathbf{x}_s,t),
\end{equation}
where $v$ denotes the subsurface velocity, $u$ represents the corresponding seismic wavefield, and $s$ is the source term. The source position is defined as $\mathbf{x}_s$, $\delta$ denotes the Dirac delta function, and $\nabla^2$ represents the Laplacian operator. The variables $\mathbf{x}$ and $t$ correspond to the spatial and temporal coordinates, respectively.

By recording the simulated wavefield $u(\mathbf{x},t)$ at receiver positions, the predicted seismic data $d_{pre}(\omega,r,t,v)$ can be obtained. This process can be written as:
\begin{equation}
\label{deqn_ex1a}
d_{pre}(r,t,v)= \int _\Omega u(\mathbf{x},t) \cdot r(\mathbf{x}_r)d\mathbf{x},
\end{equation}
where $r$ represents the receiver weighting function and $\Omega$ denotes the spatial domain where the seismic data are recorded. The receiver location is specified by $\mathbf{x}_r$. The goal of 3DFWI is to recover the subsurface velocity model $v$ by minimizing the misfit between the predicted seismic data $d_{pre}(\omega,r,t,v)$ and the observed data $d_{obs}(\omega,r,t)$.

Traditional 3DFWI formulates the objective function using the $l_2$ norm, which can be expressed as follows:
\begin{equation}
\label{deqn_ex1a}
J_{3DFWI}(v) = \dfrac{1}{2} \sum_{s} \sum_{r}\int_{t} \parallel d_{pre}(w,r,t,v) - d_{obs}(r,t)\parallel_{2}^{2} dt.
\end{equation}

The aforementioned objective function can be minimized using the gradient descent method \cite{tarantolaInversionSeismicReflection1984}, where the gradient with respect to the velocity parameter $v$ is given by:
\begin{equation}
\label{deqn_ex1a}
\dfrac{\partial J_{3DFWI}(v)}{\partial v} = \sum_{s} \sum_{r}\int_{t} [\dfrac{\partial d_{pre}(r,t,v)} {\partial v}b_{adj}(r,t,v)] dt,
\end{equation}
where $b_{adj}(r,t,v)$ denotes the adjoint source. For the $l_2$ norm objective function, it is defined as the residual between the predicted and observed seismic data $b_{adj}(r,t,v) = d_{pre}(r,t,v) - d_{obs}(r,t)$. Conventional 3DFWI directly minimizes the amplitude difference between the predicted and observed waveforms. However, when the seismic data lack low-frequency components or the initial velocity model is far from the true model, this approach often suffers from cycle skipping, which will lead to inaccurate inversion results.

\subsection{3DIFWI}\hfill

In traditional 3DFWI, the subsurface model is typically represented by a discretized velocity grid. In this method, the velocity values are directly assigned to each grid point and iteratively updated during the inversion process. However, the grid-based parameterization lacks inherent structural constraints on the model space, which reduces the stability of the inversion and makes the optimization process more susceptible to local minima.

According to the universal approximation theorem, NNs have the capability to approximate arbitrary continuous functions \cite{hornik1991approximation, leshno1993multilayer}. Based on this property, INRs have recently been introduced into FWI, leading to the development of 3DIFWI. In this framework, the velocity model is represented as a continuous function that maps spatial coordinates to velocity values, which is parameterized by a neural network. This method reparameterizes the model space from physical velocity values to the trainable parameters of the neural network. Such a parameterization implicitly imposes continuity and spatial correlation on the velocity field, rather than treating velocity values at different grid points as independent variables. In addition, it can exploit the spectral bias and implicit regularization properties of neural networks, which help guide the inversion toward smoother and more stable solutions and reduce the likelihood of convergence to local minima during the optimization process \cite{zhu2022integrating}.

In practice, 3DIFWI commonly adopts the sine representation network (SIREN) to parameterize the velocity field $v$ \cite{sitzmann2020implicit, sun2023implicit}. The SIREN architecture consists of an input layer, several hidden layers, and an output layer. In this study, we set up 4 hidden layers, each with 256 neurons. All hidden layers employ the sine activation function. We denote the SIREN output as $v'$, and add it to the initial velocity model $v_0$ to obtain the final predicted velocity $v$. This process can be expressed as follows:
\begin{equation}
\label{deqn_ex1a}
v = SIREN(\mathbf{x},\theta) \cdot \delta + v_0 = v' \cdot \delta + v_0,
\end{equation}
where $\mathbf{x}$ represents the spatial coordinates of the velocity model; $\theta$ indicates the trainable parameters in the SIREN; $v_0$ means the initial velocity for inversion.

The objective function of 3DIFWI using $l_2$ norm is as follows:
\begin{equation}
\label{deqn_ex1a}
J_{3DIFWI}(\theta) = \min_{\theta} \dfrac{1}{2} \sum_{s} \sum_{r}\int_{t} \parallel d_{pre}(r,t,\theta) - d_{obs}(r,t)\parallel_{2}^{2} dt.
\end{equation}

Optimization algorithms, such as stochastic gradient descent (SGD), limited-memory broyden-fletcher-goldfarb-shanno (LBFGS), and adaptive moment estimation (Adam) in the DL framework, can directly solve the objective function.

\subsection{The Proposed Method}\hfill

However, 3DIFWI requires all spatial coordinates of the 3D model to be input into the neural network to predict the velocity model.  For a velocity model of size $N_x\times N_y\times N_z$, the intermediate activation variables generated by inputting all spatial coordinates into the neural network are approximately $\mathcal{O}(2 \times L\times H\times N_x\times N_y\times N_z)$ in size, where $L$ is the number of hidden layers in the neural network and $H$ is the number of neurons in each layer. A typical 3D model usually contains tens of millions of grid points. Therefore, implementing 3DIFWI requires hundreds of gigabytes (GB) of graphics processing unit (GPU) or central processing unit (CPU) memory, resulting in prohibitive computational costs and a severe memory bottleneck. Furthermore, the implicit regularization inherent in the neural networks of 3DIFWI is insufficient to guarantee the structural consistency of the inversion results.

Therefore, we propose a novel tensor decomposition-based 3DIFWI framework to reduce memory requirements and improve the structural consistency of the inversion results. Tensor decomposition strategies include canonical-polyadic (CP) and tensor train (TT) decomposition. However, CP decomposition has weak expressive power and cannot capture local correlations in complex 3D velocity structures. Therefore, this proposed method selects TT decomposition as the core strategy. The TT decomposition represents a high-order tensor as a product of a series of low-rank core tensors. Adjacent core tensors are coupled through a shared rank index, forming a chain-like low-rank representation. For a velocity model $v$ of size $N_x\times N_y\times N_z$, its TT decomposition can be expressed as:
\begin{equation}
\label{TT}
v = \phi_{x} \circ \phi_{y} \circ \phi_{z},
\end{equation}
where $\phi_{x} \in \mathbb{R}^{N_x\times R_1}$,  $\phi_{y} \in \mathbb{R}^{R_1\times N_y\times R_2}$,  and $\phi_{z} \in \mathbb{R}^{R_2\times N_z}$ represent the low-rank core tensors of TT corresponding to the three spatial dimensions $x$, $y$, and $z$, respectively. $R_1$ and $R_2$ denotethe ranks of the TT decomposition, used to control the degree of low rank in the decomposition. We set both $R_1$ and $R_2$ to 64. As an intermediate core tensor, $\phi_{y}$ is coupled with adjacent modes through shared rank indices $R_1$ and $R_2$, realizing low-rank association and information transmission between different spatial dimensions.

Then, we construct three axis-specific neural networks using the SIREN network  to predict the low-rank core tensors $\phi_{x}$, $\phi_{y}$, and $\phi_{z}$ of TT decomposition, as follows:
\begin{equation}
\phi_{x} = SIREN_x(c_x,\theta_x),
\end{equation}
\begin{equation}
\phi_{y} = SIREN_y(c_y,\theta_y),
\end{equation}
\begin{equation}
\phi_{z} = SIREN_z(c_z,\theta_z),
\end{equation}
where $SIREN_x$, $SIREN_y$ and $SIREN_z$ represent axis-specific subnetworks constructed for the $x$, $y$ and $z$ axes directions, respectively. $c_x \in \mathbb{R}^{N_x\times 1}$, $c_y  \mathbb{R}^{N_y\times 1}$, and $c_z  \mathbb{R}^{N_z\times 1}$ are the one-dimensional spatial coordinate inputs on the corresponding coordinate axes. $\theta_x$, $\theta_y$, and $\theta_z$ denote the trainable parameters of the subnetworks.

Since each axis-specific subnetwork only processes one-dimensional vector coordinate inputs, the size of the intermediate activation variables generated by TT-3DIFWI is approximately $\mathcal{O}(2 \times L\times H\times (N_x+ N_y+ N_z))$. Compared to $\mathcal{O}(2 \times L\times H\times N_x\times N_y\times N_z)$ in 3DIFWI, this significantly reduces the computer memory required for training. Furthermore, due to the low-rank property of TT decomposition, TT-3DIFWI can guarantee the structural consistency of the inversion result.

\begin{figure}
\centering
\includegraphics[width=3.5in]{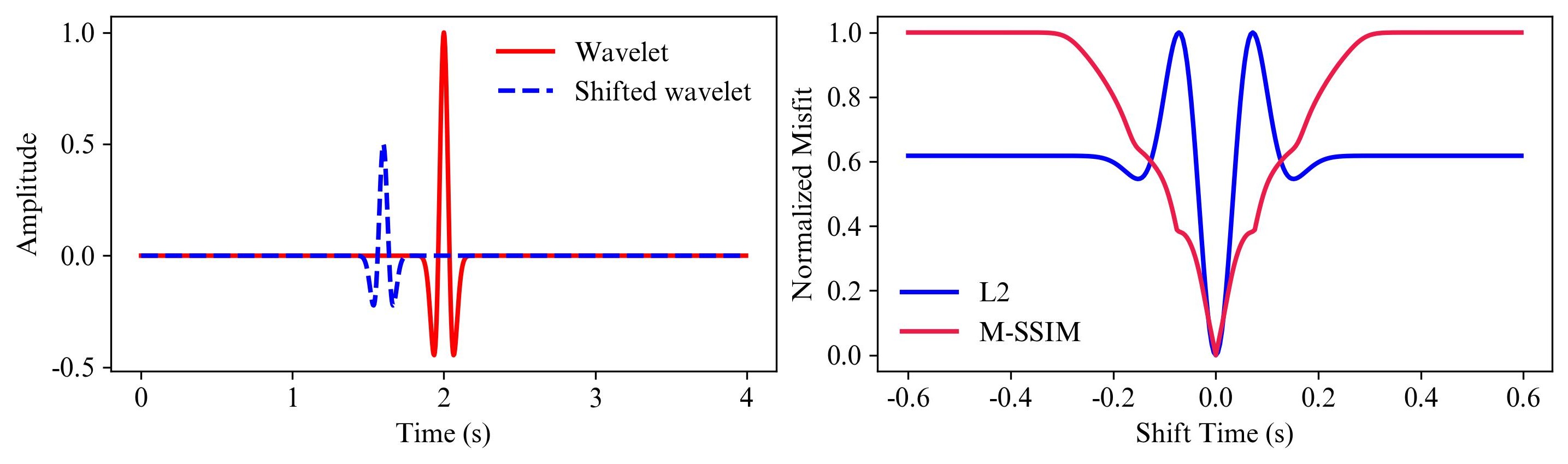}
\caption{Ricker wavelet misfit function test. (a) Ricker wavelet. (b) Nominalization misfit.}
\label{./misfit.jpg}
\end{figure}

Furthermore, the multi-scale structural similarity index measure (M-SSIM) is employed to quantify the multi-scale structural differences between the observed and predicted seismic data \cite{he2025ms_atpv}. M-SSIM enables robust comparison across multiple scales by progressively capturing both large-scale and fine-scale structural similarities. This is achieved by introducing multi-scale Gaussian filtering with different Gaussian window standard deviations ($\sigma_w$) in both the time and spatial dimensions. The M-SSIM can be formulated as follows:
\begin{equation}
\label{deqn_ex1a}
MS(d_{pre} , d_{obs}) = \sum_{i=1}^n L (d_{pre(i)} , d_{obs(i)}) \cdot C (d_{pre(i)} , d_{obs(i)}) \cdot S (d_{pre(i)} , d_{obs(i)}) ,
\end{equation}
where $L$, $C$, and $S$ denote the local mean amplitude, local energy, and local waveform contrast terms, respectively. The index $i$ represents the $i$-th Gaussian window scale, and the symbol $\cdot$ denotes the dot product operation.

The local mean amplitude similarity term $L$ measures the correlation between the Gaussian-smoothed predicted and observed seismic data. This term characterizes the similarity of local mean amplitudes and enhances data fitting at local scales, thereby helping mitigate the cycle skipping problem \cite{zhang2019local, huang2023toward}. At the $i$-th Gaussian window scale, the term $L$ is defined as:
\begin{equation}
\label{L}
L(d_{pre(i)} , d_{obs(i)}) = \dfrac{2 \mu_{d_{pre(i)}} \cdot \mu_{d_{obs(i)}}+C_1}{\mu_{d_{pre(i)}}^2 + \mu_{d_{obs(i)}}^2 +C_1},
\end{equation}
where $\mu_{d_{pre(i)}} = G_{\sigma_{w(i)}} \ast d_{pre}$ and $\mu_{d_{obs(i)}} = G_{\sigma_{w(i)}} \ast d_{obs}$ denote the local mean amplitudes of the predicted and observed data within the Gaussian window $G_{\sigma_{w(i)}}$, respectively. Here, $\ast$ represents the convolution operation, and $C_1 = 1 \times 10^{-4}$ is a small constant introduced to ensure numerical stability. When the local mean amplitudes of the predicted and observed data are similar, the value of $L$ approaches 1, indicating a high degree of structural similarity. In contrast, large amplitude differences lead to smaller values of $L$, reflecting reduced structural similarity.

Similarly, the local energy similarity term $C$ measures the correlation between the local energies of the predicted and observed seismic data. Because local energy is dominated by low-frequency components, it provides a more stable measure for waveform comparison and helps mitigate the cycle skipping problem \cite{liu2018full, li2025addressing}. The computation of $C$ at the $i$-th Gaussian window scale is defined as:
\begin{equation}
\label{C}
C(d_{pre(i)} , d_{obs(i)}) = \dfrac{2 \sigma_{d_{pre(i)}} \cdot \sigma_{d_{obs(i)}} + C_2}{{\sigma_{d_{pre(i)}}}^2 + {\sigma_{d_{obs(i)}}}^2 +C_2},
\end{equation}
where $\sigma_{d_{pre(i)}} = \bigl(G_{\sigma_{w(i)}} \ast d_{pre}^2 - \mu_{d_{pre(i)}}^2\bigr)^{1/2}$ and $\sigma_{d_{obs(i)}} = \bigl(G_{\sigma_{w(i)}} \ast d_{obs}^2 - \mu_{d_{obs(i)}}^2\bigr)^{1/2}$ denote the local standard deviations representing the local energies of the predicted and observed data, respectively. Here, $G_{\sigma_{w(i)}}$ represents the Gaussian window with standard deviation $\sigma_{w(i)}$, and $C_2 = 9 \times 10^{-4}$ is a small constant introduced to ensure numerical stability. When the local energy distributions of the predicted and observed data are similar, the value of $C$ approaches 1, indicating a high degree of similarity. Conversely, large differences in local energies lead to smaller values of $C$, reflecting reduced similarity between the two datasets.

Finally, the local waveform similarity term $S$ measures the covariance between the predicted and observed seismic data, thereby characterizing the consistency of local waveform structures. The computation of $S$ at the $i$-th Gaussian window scale is defined as:
\begin{equation}
\label{S}
S(d_{pre(i)} , d_{obs(i)}) = \dfrac{\sigma_{d_{pre(i)} \cdot d_{obs(i)}}+C_3}{\sigma_{d_{pre(i)}} \cdot \sigma_{d_{obs(i)}} + C_3},
\end{equation}
where $\sigma_{d_{pre(i)} d_{obs(i)}} = G_{\sigma_{w(i)}} \ast (d_{pre} \cdot d_{obs}) - \mu_{d_{pre(i)}} \mu_{d_{obs(i)}}$ denotes the local covariance between the predicted and observed data. This covariance term quantifies the similarity of local waveform variations between the two datasets. Here, $C_3 = 4.5 \times 10^{-4}$ is a small constant introduced to ensure numerical stability. When the local waveform patterns of the predicted and observed data are highly consistent, the value of $S$ approaches 1, indicating strong structural similarity. Conversely, larger waveform discrepancies result in smaller values of $S$, reflecting reduced similarity between the two datasets.

\begin{figure}
\centering
\includegraphics[width=3.5in]{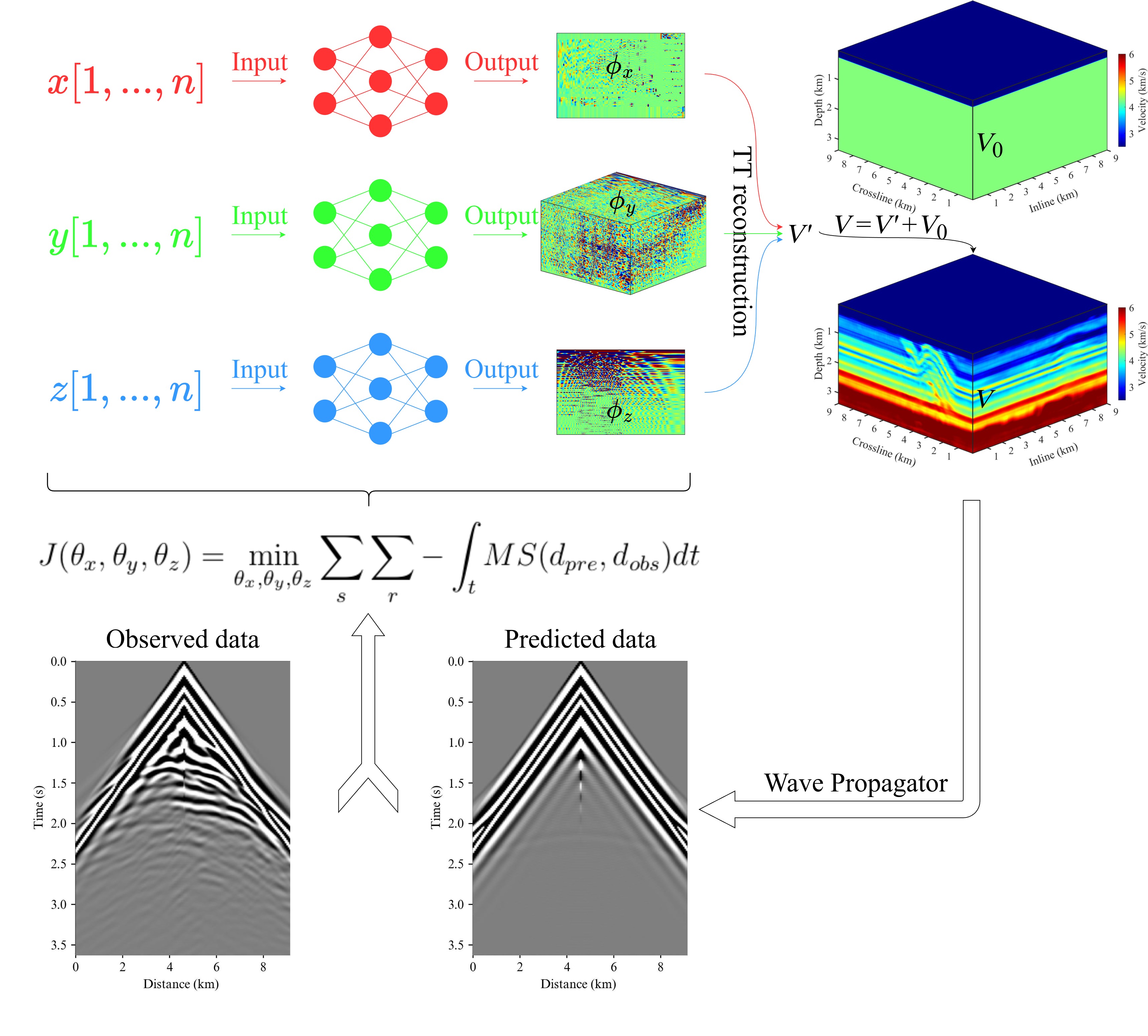}
\caption{Workflow of the proposed method.}
\label{./LR_IFWI.jpg}
\end{figure}

The combination of the three similarity terms effectively captures the multi-scale structural discrepancies between the predicted and observed seismic data. To extract structural features at different scales while maintaining computational efficiency, the Gaussian window standard deviations ($\sigma_w$) in the M-SSIM objective function are set to 0.5, 1, and 2. To further illustrate the advantages of M-SSIM, a synthetic example using a Ricker wavelet and a time-shifted Ricker wavelet is designed, as shown in Figure \ref{./misfit.jpg}(a). Figure \ref{./misfit.jpg}(b) presents a comparison of the normalized misfit curves obtained using M-SSIM and the conventional $l_2$ norm. The results demonstrate that the $l_2$ norm is prone to local minima caused by cycle skipping, whereas M-SSIM effectively avoids local minima and converges to the global minimum.

Finally, the objective function of the proposed method is:
\begin{equation}
\label{objective function}
J(\theta_x,\theta_y,\theta_z) = \min_{\theta_x,\theta_y,\theta_z} \sum_{s} \sum_{r} -\int_{t} MS(d_{pre} , d_{obs})dt.
\end{equation}

The workflow of the proposed method is shown in Figure \ref{./LR_IFWI.jpg}. The overall inversion framework of the proposed method is shown in Algorithm 1.

\begin{algorithm}
\caption{The proposed method algorithm.}
\textbf{Input:} $v_0$: initial velocity model; $c_x$, $c_y$, and $c_z$: vector coordinates; $SIREN_x$, $SIREN_y$, and $SIREN_z$: axis-specific sine representation networks; $n_s$: number of shot gathers; $d_{obs}^{n_s}$: observed data; $n_w$: number of gaussian windows of different scales; $\sigma_w^{n_w}$: gaussian window sigma value; $lr$: learning rate for updating velocity model; $s$: source wavelet; $B$: number of batches; $E$: number of epochs.\\
\textbf{Output:} inverted velocity model $v$.
\begin{algorithmic}[1]
\State Initialize $SIREN_x$, $SIREN_y$, and $SIREN_z$ trainable parameters $\theta_x$, $\theta_y$, and $\theta_z$.
\State Calculate velocity model using TT decomposition $v = (SIREN_x(c_x,\theta_x) \circ SIREN_y(c_y,\theta_y) \circ SIREN_z(c_z,\theta_z)) + v_0 =  (\phi_{x} \circ \phi_{y} \circ \phi_{z}) + v_0 = v' + v_0$.
\For{$epoch = 1,\dots, E$}
  \For{$b = 1,\dots, B$}
    \State Sample data $d_{obs}^{c}$ with $c = n_s / B$ from observed data  $d_{obs}^{n_s}$.
    \State \parbox[t]{0.95\linewidth}{%
    \hangindent=0em \hangafter=1
    Generate the corresponding predicted data $d_{pre}^{c} $ from current velocity model $v$.}
 
    \For{$i = 1,\dots, n_w$}

    \State \parbox[t]{0.95\linewidth}{%
    \hangindent=0em \hangafter=1
    Extract the mean ($\mu_{d_{pre(i)}}^{c}$, $\mu_{d_{obs(i)}}^{c}$), standard deviation ($\sigma_{d_{pre(i)}}^{c}$, $\sigma_{d_{obs(i)}}^{c}$), and covariance ($\sigma_{d_{pre(i)}}^{c} \cdot \sigma_{d_{obs(i)}}^{c}$, $\sigma_{d_{pre(i)} \cdot d_{obs(i)}}^{c}$) features of the predicted $d_{pre}^{c}$ and observed $d_{obs}^{c}$ data under the $i$-th scale window corresponding to the $\sigma_w^i$ value.}
    \State \parbox[t]{0.95\linewidth}{%
    \hangindent=0em \hangafter=1
    Calculate the correlation of the local mean amplitude $\mu_{d_{pre(i)}}^{c}$ and $\mu_{d_{obs(i)}}^{c}$ using equation (\ref{L}).}
    
    \State \parbox[t]{0.95\linewidth}{%
    \hangindent=0em \hangafter=1
    Calculate the correlation of the local energies $\sigma_{d_{pre(i)}}^{c}$ and $\sigma_{d_{obs(i)}}^{c}$ using equation (\ref{C}).}
    
    \State \parbox[t]{0.95\linewidth}{%
    \hangindent=0em \hangafter=1
    Calculate the local waveform similarity through $\sigma_{d_{pre(i)}}^{c} \cdot \sigma_{d_{obs(i)}}^{c}$ and $ \sigma_{d_{pc(i)} \cdot d_{oc(i)}}^{c}$ using equation (\ref{S}).}
    \EndFor

    \State Calculate the loss using equation (\ref{objective function}).
    \State \parbox[t]{0.95\linewidth}{%
    \hangindent=0em \hangafter=1
    Update $\theta_x$, $\theta_y$, and $\theta_z$ using the adam optimizer with learning rate $lr$.}
    \State \parbox[t]{0.95\linewidth}{%
    \hangindent=0em \hangafter=1
    Calculate velocity model using TT decomposition $v = (SIREN_x(c_x,\theta_x) \circ SIREN_y(c_y,\theta_y) \circ SIREN_z(c_z,\theta_z)) + v_0 =  (\phi_{x} \circ \phi_{y} \circ \phi_{z}) + v_0 = v' + v_0$.}
  \EndFor
\EndFor
\State \Return final inverted velocity model $v$.
\end{algorithmic}
\end{algorithm}

\begin{figure}
\centering
\includegraphics[width=3.5in]{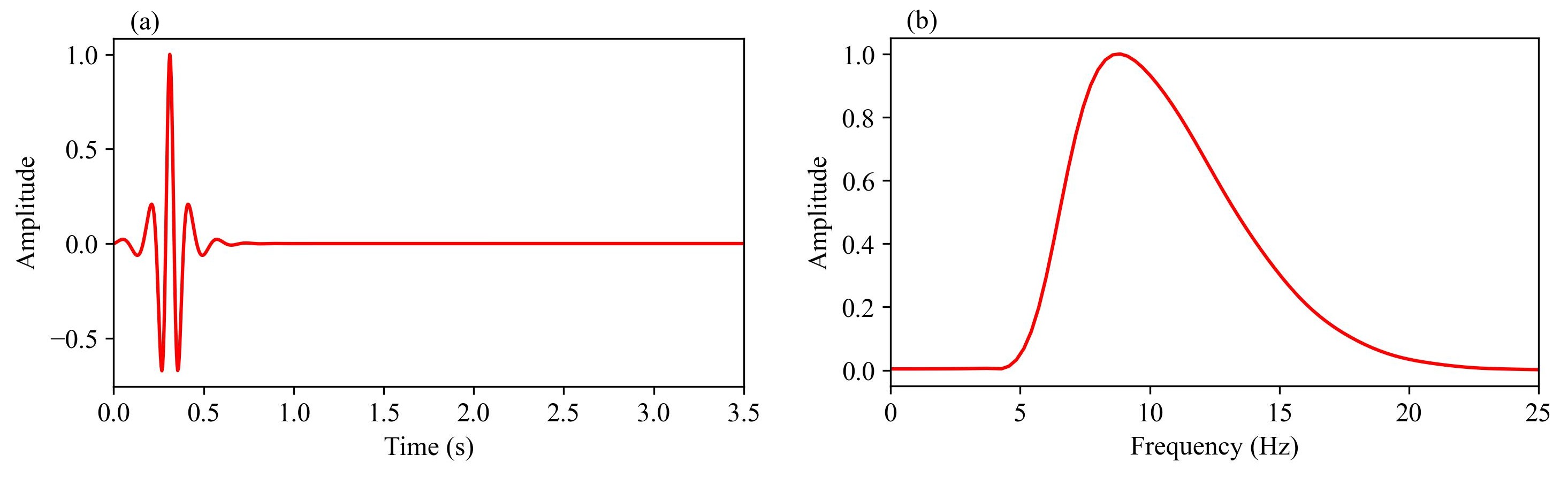}
\caption{The source wavelet and corresponding frequency spectrum. (a) Source wavelet. (b) Frequency spectrum.}
\label{./source.jpg}
\end{figure}

\begin{figure}
\centering
\includegraphics[width=1.5in]{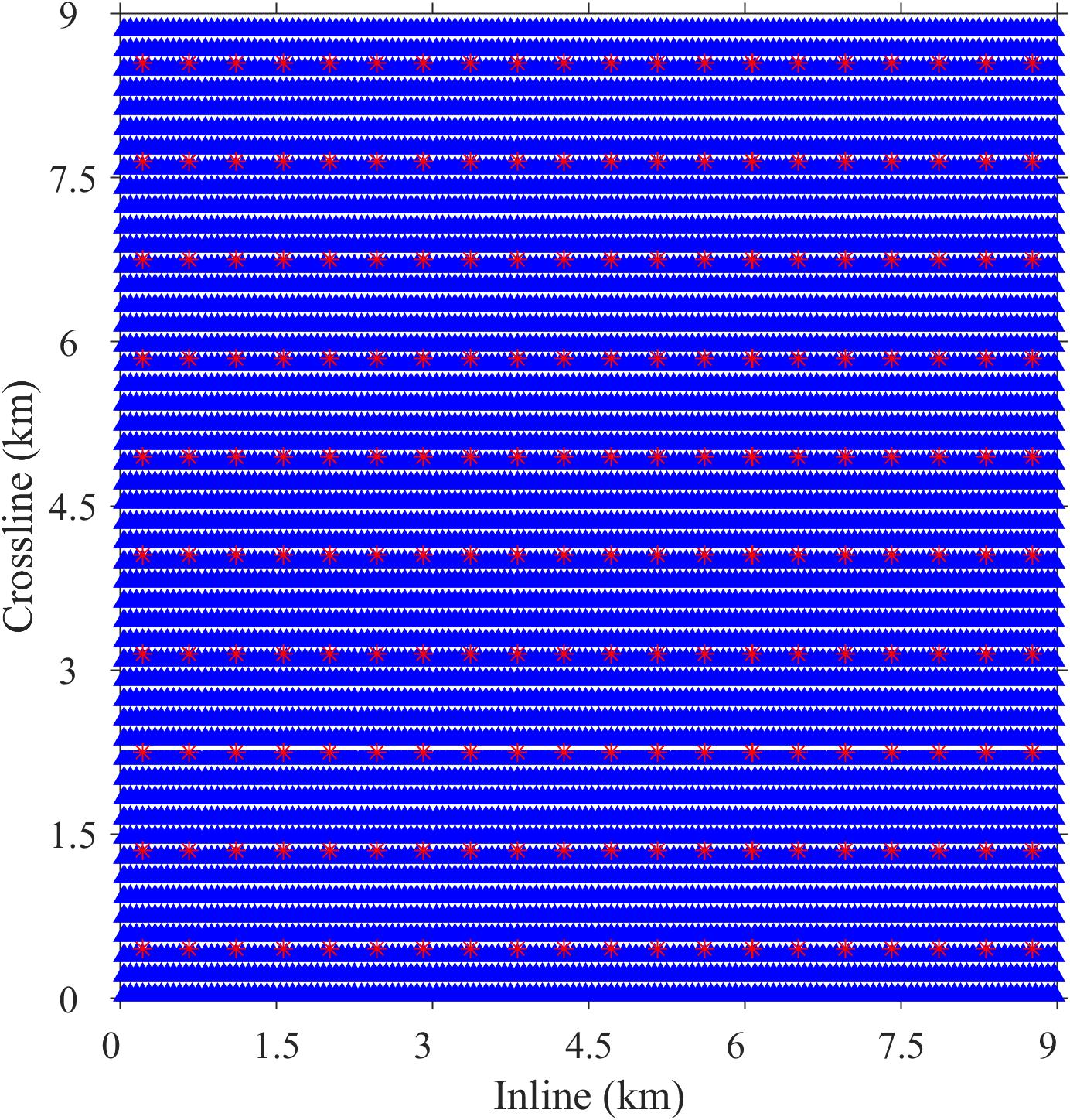}
\caption{The synthetic data acquisition geometry.}
\label{./overthrust/sr.jpg}
\end{figure}

\begin{figure}
\centering
\includegraphics[width=2in]{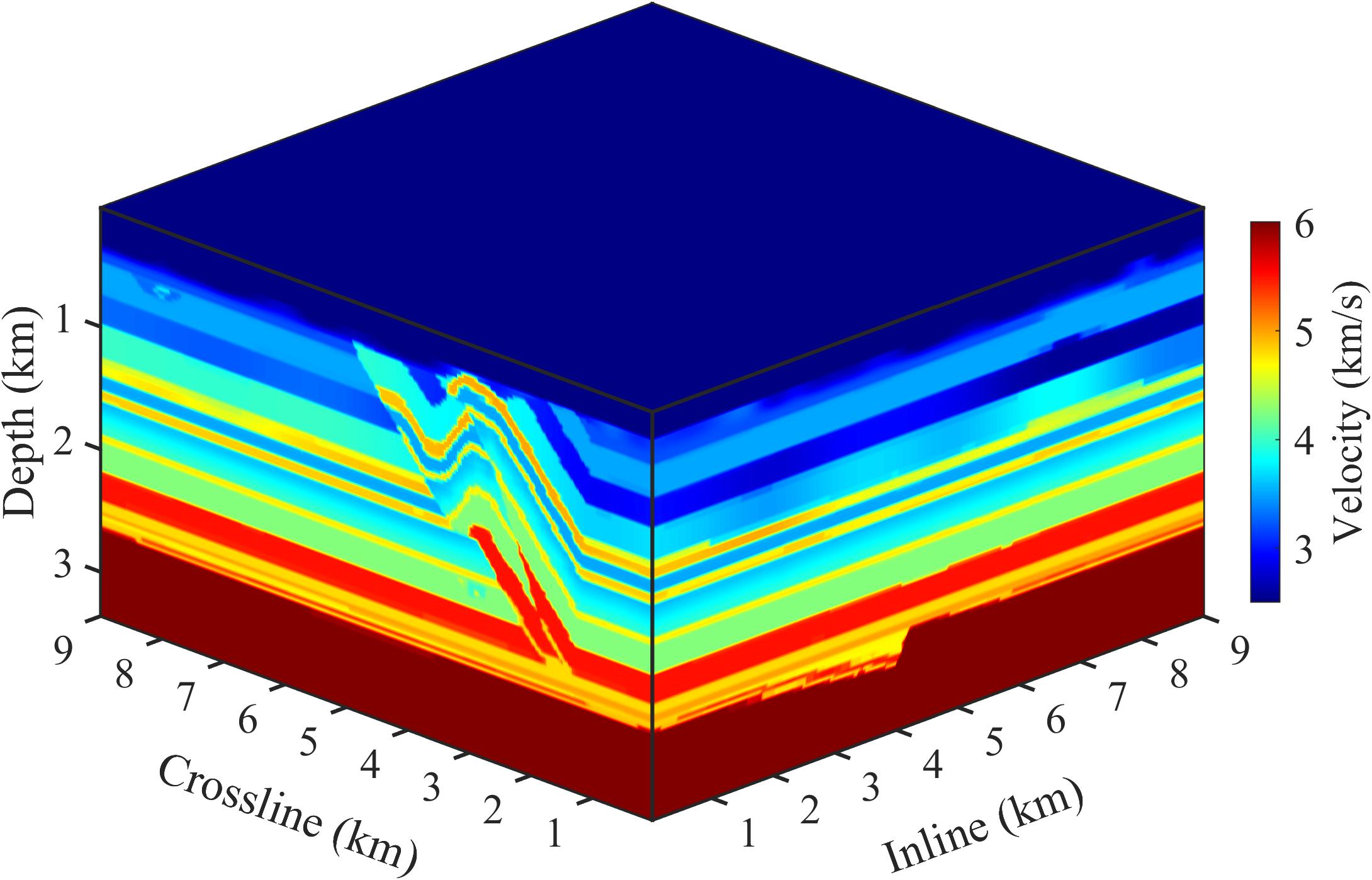}
\caption{Overthrust velocity model.}
\label{./overthrust/True1.jpg}
\end{figure}

\section{Experiments and Results}\hfill

This section compares and evaluates the proposed method with alternative methods using synthetic and field datasets. In the synthetic dataset, the dominant frequency of the source wavelet used to generate the synthetic data is 8 Hz, and a high-pass filter is used to remove low-frequency components below 5 Hz, as shown in Figure \ref{./source.jpg}. In the field dataset, seismic data is affected by noise and surface wave interference, and low-frequency components below 5 Hz are destroyed. For synthetic data, we tested using a constant gradient velocity model and a uniform velocity model as the initial models for inversion, respectively. For field data, we use a constant gradient velocity model as the initial model for inversion. We set the total number of inversion epochs to 200. Detailed experimental settings are as follows.

\subsection{3D Overthrust Model}\hfill

\begin{figure*}
\centering
\includegraphics[width=6.5in]{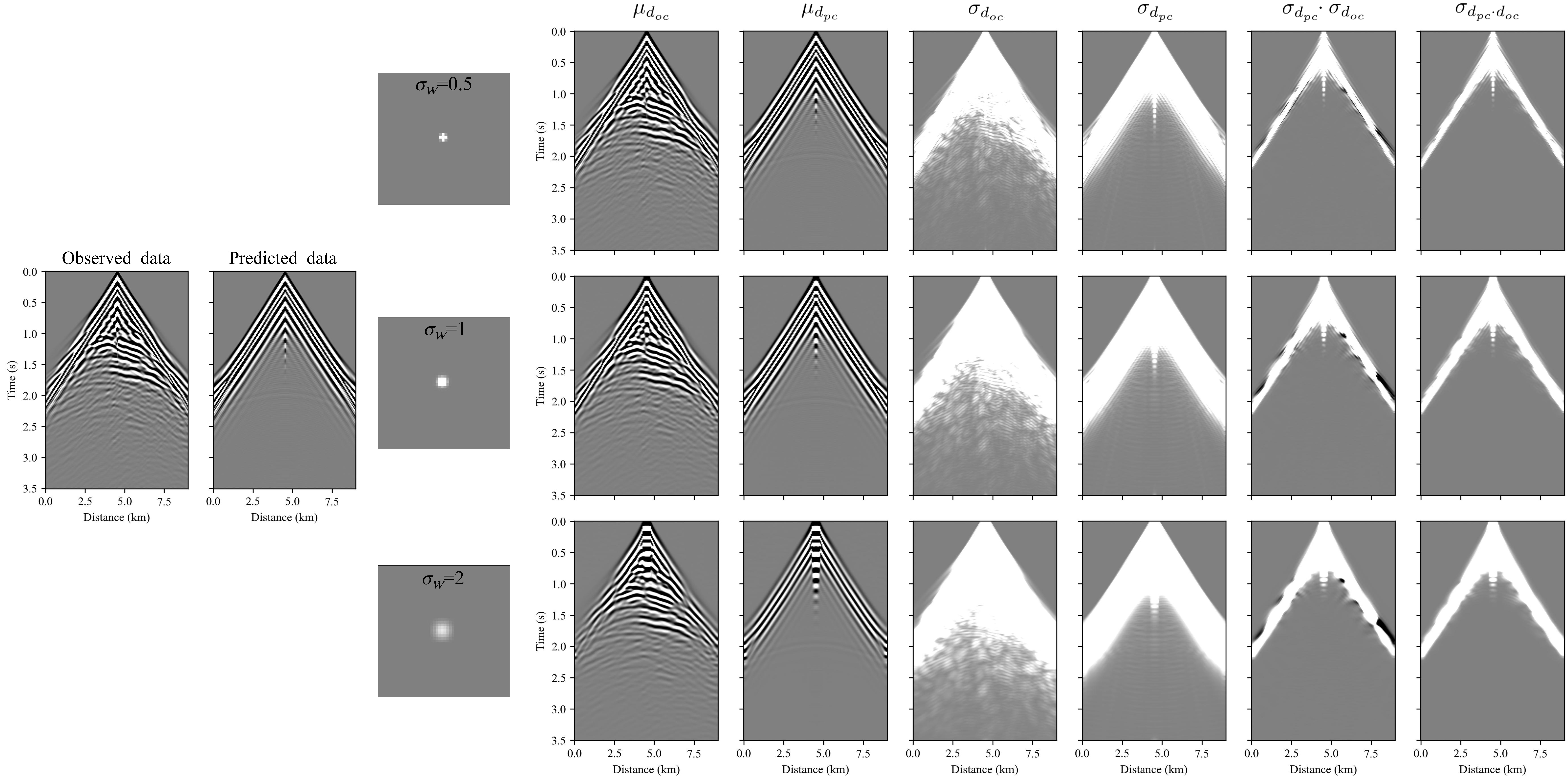}
\caption{M-SSIM extracts multi-scale structural features from predicted and observed data.}
\label{./overthrust/SSIM_data.jpg}
\end{figure*}

\begin{figure}
\centering
\includegraphics[width=3.5in]{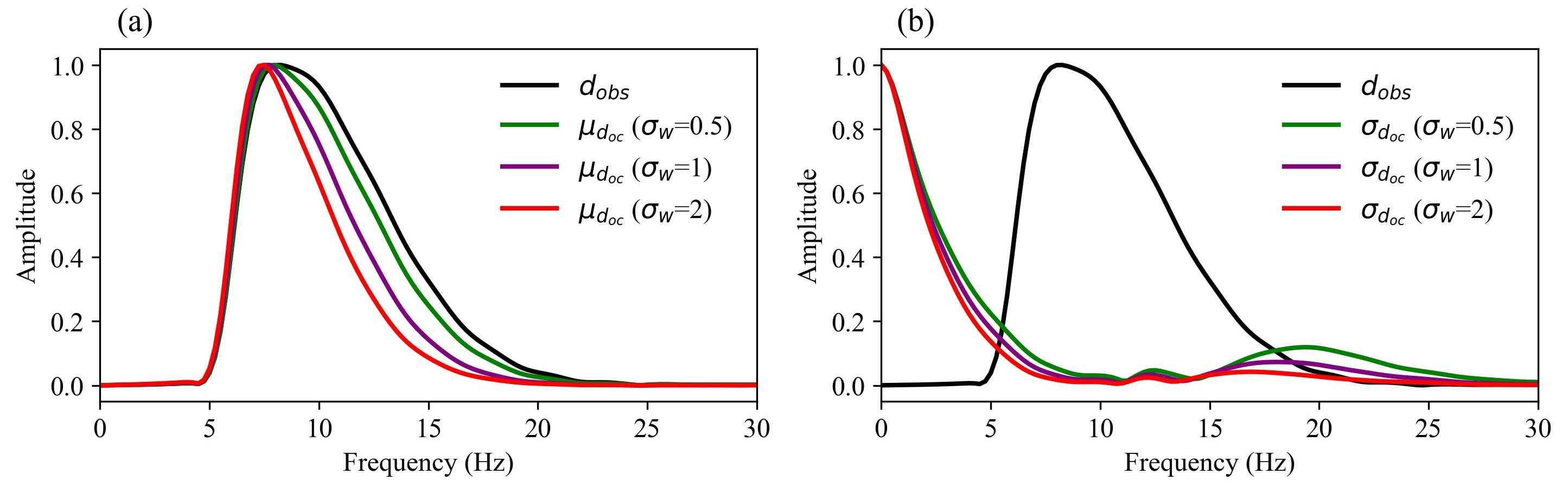}
\caption{Normalized frequency spectrum of multi-scale structural features. (a) Local mean amplitude. (b) Local energy.}
\label{./overthrust/mean_variance.jpg}
\end{figure}

\begin{figure*}
\centering
\includegraphics[width=6in]{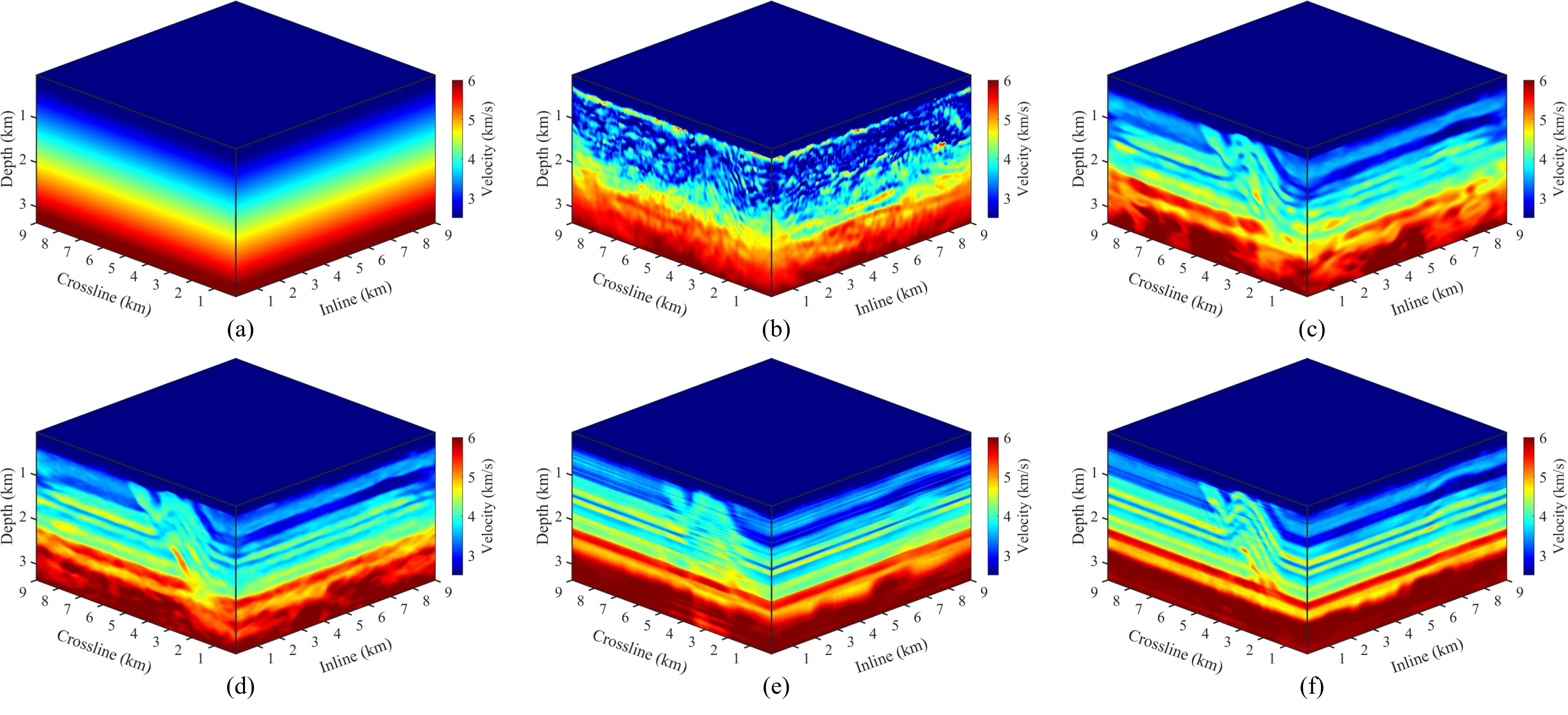}
\caption{Constant gradient initial model and inversion results. (a) Initial modeld. (b) 3DFWI. (c) 3DIFWI. (d) 3DIFWI with M-SSIM. (e) CP-3DIFWI with M-SSIM. (f) TT-3DIFWI with M-SSIM.}
\label{./overthrust/result.jpg}
\end{figure*}

\begin{figure*}
\centering
\includegraphics[width=6in]{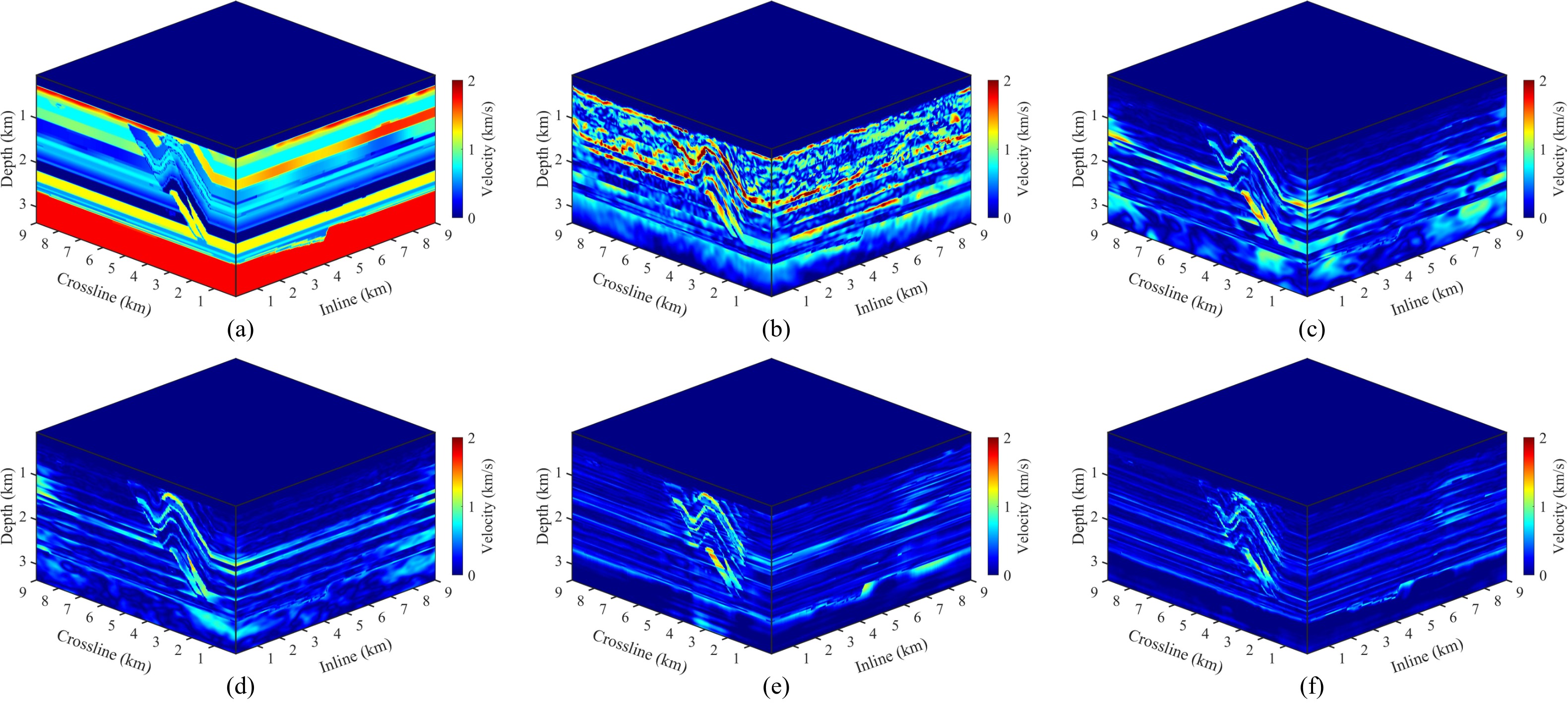}
\caption{The absolute error between the inversion results and the true model. (a) Initial modeld. (b) 3DFWI. (c) 3DIFWI. (d) 3DIFWI with M-SSIM. (e) CP-3DIFWI with M-SSIM. (f) TT-3DIFWI with M-SSIM.}
\label{./overthrust/resultr.jpg}
\end{figure*}

\begin{figure}
\centering
\includegraphics[width=3.5in]{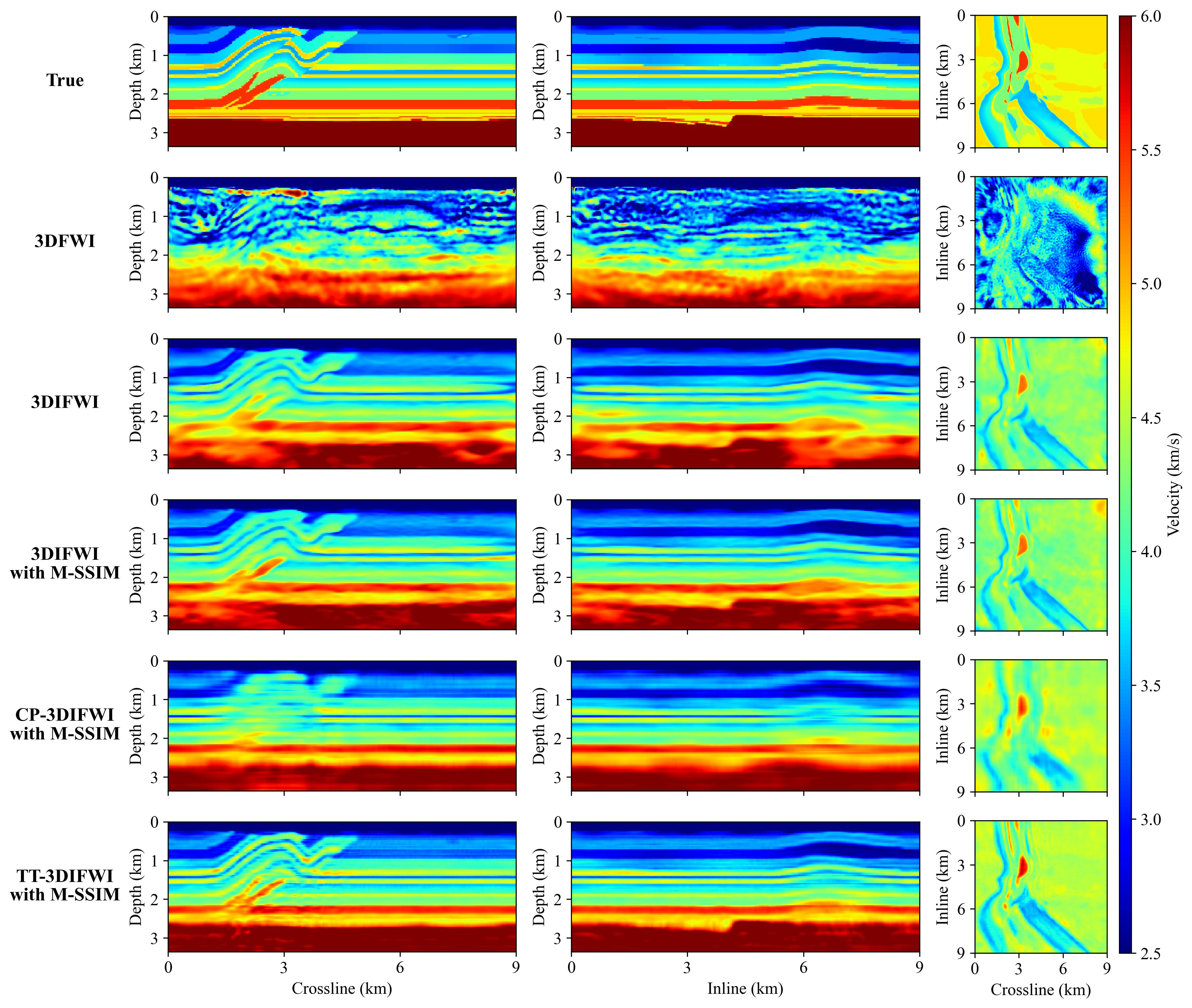}
\caption{The slices results obtained from constant gradient initialization model at Inline = 0.6 km, crossline = 0.3 km, and depth = 1.5 km.}
\label{./overthrust/section_result.jpg}
\end{figure}

\begin{figure}
\centering
\includegraphics[width=3.5in]{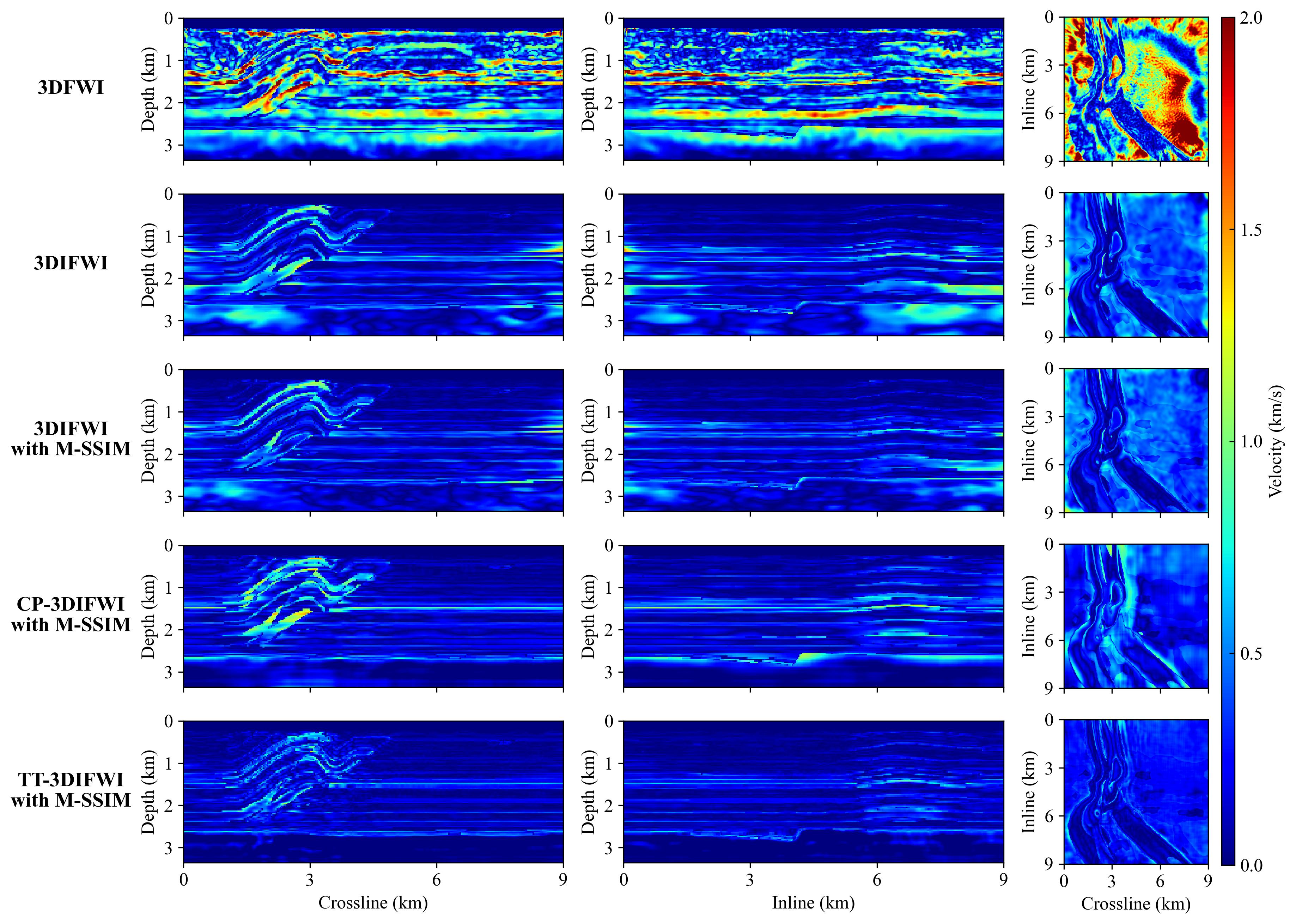}
\caption{The slices absolute error obtained from constant gradient initialization model at Inline = 0.6 km, crossline = 0.3 km, and depth = 1.5 km.}
\label{./overthrust/section_resultr.jpg}
\end{figure}

\begin{figure*}
\centering
\includegraphics[width=6in]{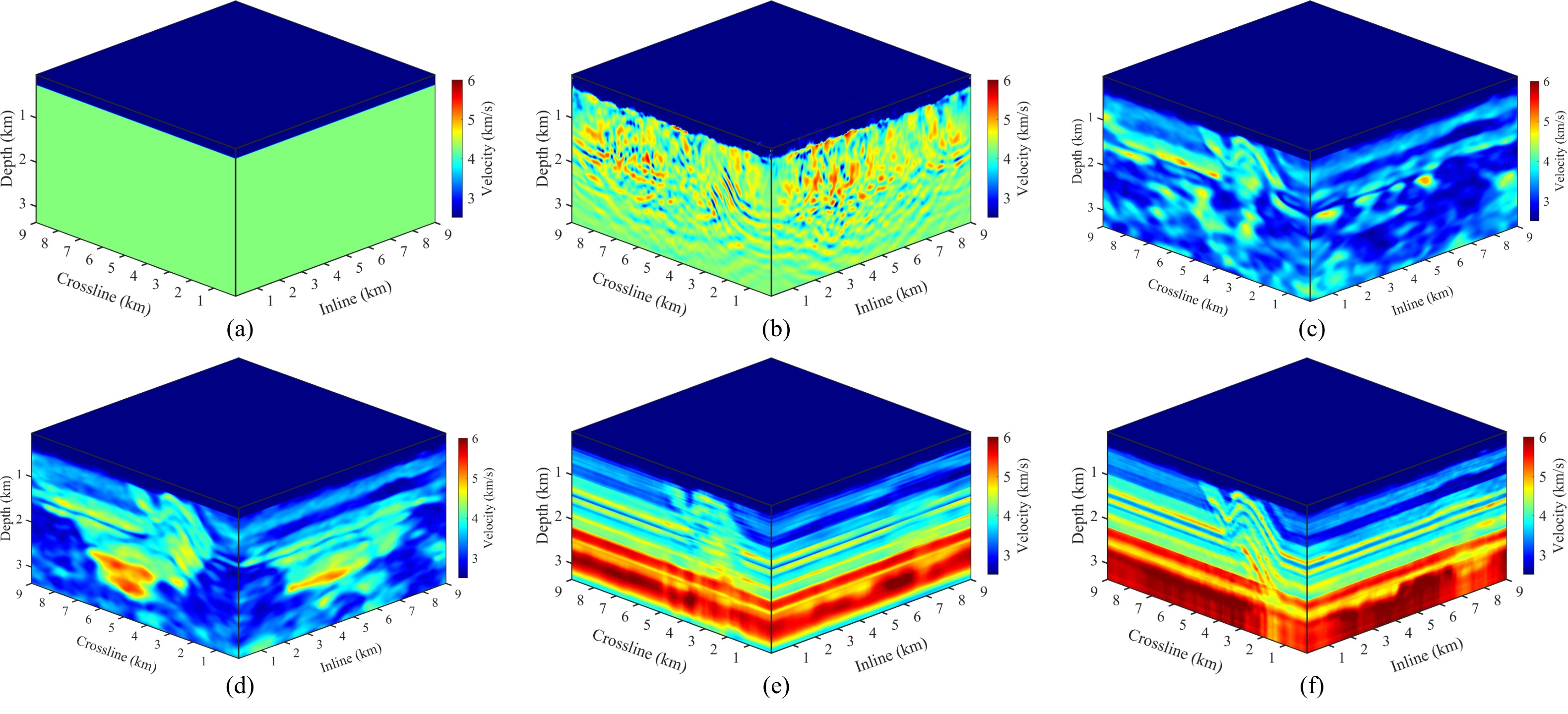}
\caption{Homogeneous gradient initial model and inversion results. (a) Initial modeld. (b) 3DFWI. (c) 3DIFWI. (d) 3DIFWI with M-SSIM. (e) CP-3DIFWI with M-SSIM. (f) TT-3DIFWI with M-SSIM.}
\label{./overthrust/result_homo.jpg}
\end{figure*}

\begin{figure}
\centering
\includegraphics[width=3.5in]{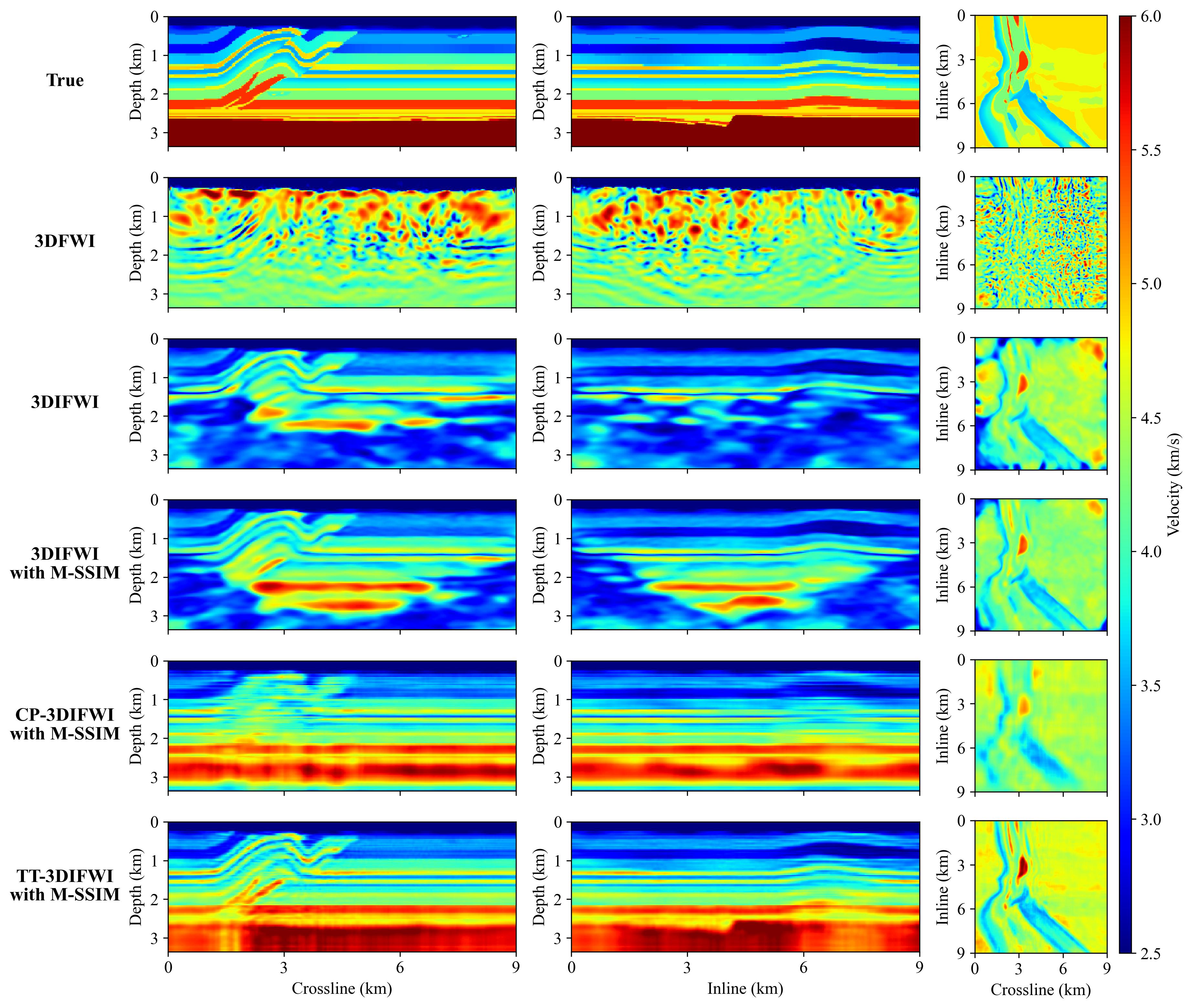}
\caption{The slices results obtained from homogeneous gradient initialization model at Inline = 0.6 km, crossline = 0.3 km, and depth = 1.5 km.}
\label{./overthrust/section_result_homo.jpg}
\end{figure}

\begin{figure*}
\centering
\includegraphics[width=6in]{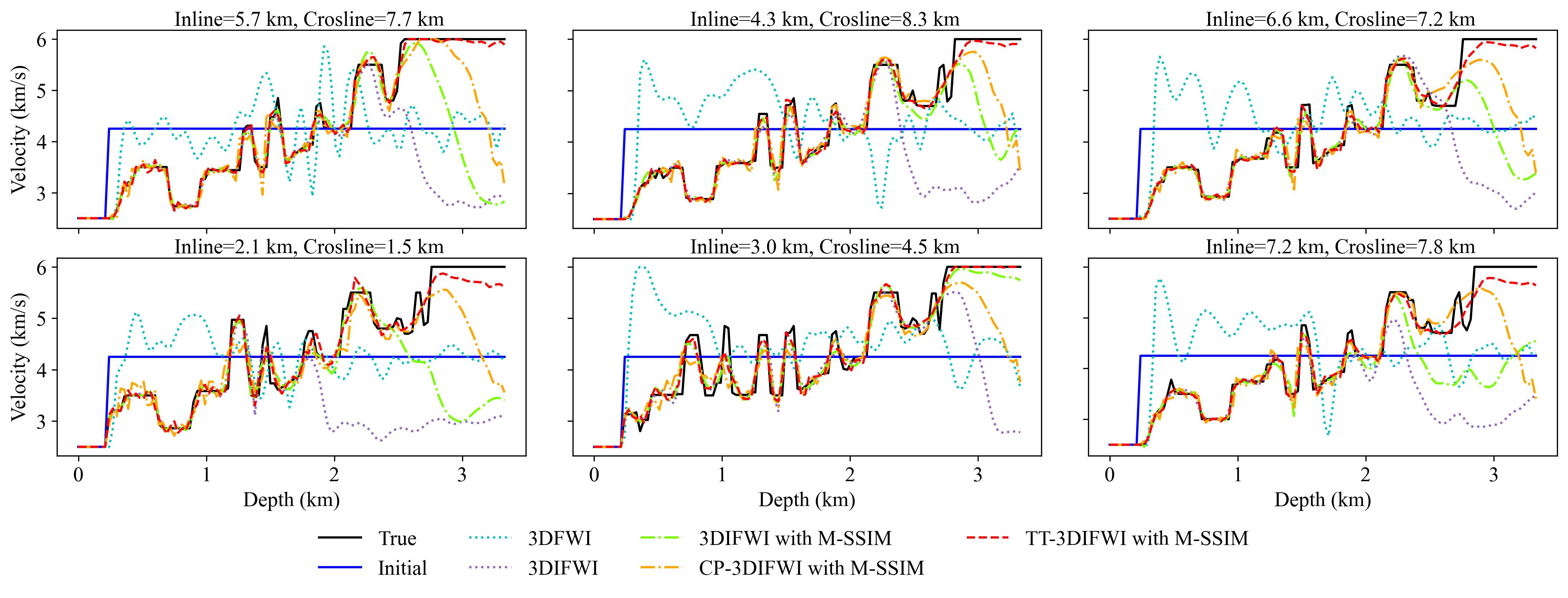}
\caption{Comparison of velocity curves.}
\label{./overthrust/Single_homo.jpg}
\end{figure*}

First, we use the 3D overthrust model with a spatial grid size of $112\times300\times300$ and a spatial sampling interval of 30 m to test the feasibility of the proposed method. We arrange 20 sources at equal intervals along the model surface in the inline direction and 10 lines of sources in the crossline direction. Similarly, we set up 150 receivers at equal intervals along the inline direction and 100 lines of receivers in the crossline direction. The seismic data temporal interval is 4 ms, and the length of the record is 3.5 s. The data acquisition geometry is shown in Figure \ref{./overthrust/sr.jpg}, where red stars represent sources and blue triangles are receivers. Figure \ref{./overthrust/SSIM_data.jpg} presents the local mean amplitude, local energy, and local covariance of the predicted and observed data for different values of $\sigma_w$, as computed using the M-SSIM method. These multi-scale structural features are then utilized to calculate similarity measures via equations (\ref{L}), (\ref{C}), and (\ref{S}), which subsequently guide the model updates. Figure \ref{./overthrust/mean_variance.jpg} compares the normalized frequency spectrum of the observed data and multi-scale structural features. As the $\sigma_w$ increases, the low-frequency components of the local average amplitude and local energy data increase. This helps reduce the risk of cycle skipping during the inversion process.

Figure \ref{./overthrust/True1.jpg}a shows the true 3D Overthrust model, and Figure \ref{./overthrust/True1.jpg}b is the constant gradient initial model. The inversion results obtained using 3DFWI, 3DIFWI, 3DIFWI with M-SSIM, CP-3DIFWI with M-SSIM, and TT-3DIFWI with M-SSIM are presented in Figures \ref{./overthrust/result.jpg}a–\ref{./overthrust/result.jpg}f, respectively. The conventional 3DFWI result (Figure \ref{./overthrust/result.jpg}b) exhibits poor accuracy and structural continuity due to severe cycle skipping, failing to recover the correct subsurface velocity structures. By contrast, 3DIFWI (Figure \ref{./overthrust/result.jpg}c) employs INR to reparameterize the velocity model. Benefiting from the spectral bias property of INR, the inversion result shows improved structural continuity and overall accuracy. When the M-SSIM objective function is incorporated (Figure \ref{./overthrust/result.jpg}d), the inversion further benefits from multi-scale structural similarity constraints between predicted and observed data. This strategy alleviates cycle skipping and enhances the capability to recover complex geological structures. For the tensor decomposition–based approaches, CP-3DIFWI with M-SSIM (Figure \ref{./overthrust/result.jpg}e) reconstructs the velocity model by predicting factor matrices through axis-specific neural networks and combining them via tensor operations. Although this approach improves the structural consistency of the inversion result, the limited representation capability of CP decomposition restricts its ability to capture fine-scale and complex structural details. In comparison, TT-3DIFWI with M-SSIM (Figure \ref{./overthrust/result.jpg}f) introduces tensor train decomposition, which possesses stronger representation capability due to the involvement of core tensors. As a result, this method achieves more accurate reconstruction of both large-scale velocity structures and small-scale geological features, leading to the most reliable inversion result among all tested methods.

The absolute error between the inversion results and the true velocity model is calculated and presented in Figure \ref{./overthrust/resultr.jpg}. We can see that, the conventional 3DFWI result exhibits the largest error, indicating its limited capability in accurately recovering subsurface velocity structures under the influence of cycle skipping. Compared with 3DFWI, 3DIFWI significantly reduces the inversion error by employing INR to reparameterize the velocity model. When the M-SSIM objective function is incorporated, the inversion error is further reduced, demonstrating that multi-scale structural similarity constraints can effectively improve inversion accuracy. Moreover, the tensor decomposition–based approaches, namely CP-3DIFWI with M-SSIM and TT-3DIFWI with M-SSIM, further decrease the overall inversion error. Among these methods, TT-3DIFWI with M-SSIM achieves the lowest error, indicating its superior capability in representing complex velocity structures. These results suggest that reconstructing the velocity model using tensor decomposition enhances the structural consistency of the inversion results and improves overall inversion accuracy.

To further evaluate the inversion performance, orthogonal slices along the inline, crossline, and depth directions are compared for different methods, as shown in Figure \ref{./overthrust/section_result.jpg}. The rows from top to bottom correspond to the true model, 3DFWI, 3DIFWI, 3DIFWI with M-SSIM, CP-3DIFWI with M-SSIM, and TT-3DIFWI with M-SSIM, respectively. The slices obtained from conventional 3DFWI exhibit severe cycle-skipping artifacts, resulting in distorted structures and inaccurate velocity distributions. In comparison, both 3DIFWI and 3DIFWI with M-SSIM effectively mitigate cycle skipping. However, the recovered velocity structures still exhibit limited structural consistency, and the stratigraphic boundaries remain relatively blurred. In contrast, the tensor decomposition–based methods, namely CP-3DIFWI with M-SSIM and TT-3DIFWI with M-SSIM, produce slice results with improved structural continuity and clearer stratigraphic interfaces. Nevertheless, due to the limited representation capability of CP decomposition, CP-3DIFWI with M-SSIM shows reduced ability to characterize complex geological features. Benefiting from the higher expressive power provided by the core tensor structure, TT-3DIFWI with M-SSIM achieves more accurate reconstruction of complex velocity structures. Figure \ref{./overthrust/section_resultr.jpg} shows the absolute error between the slice inversion results and the true model. Compared to 3DFWI, standard 3DIFWI, and M-SSIM-based 3DIFWI, the results of tensor decomposition-based TT-3DIFWI with M-SSIM have significantly smaller absolute errors. However, the CP-3DIFWI with M-SSIM results show larger errors in structurally complex regions.

In addition, we calculate the total GPU memory required for forward and backward propagation of INR in the proposed method and other methods, as shown in Table \ref{tab:gpu_memory}. It is worth noting that the results in Table \ref{tab:gpu_memory} do not include the memory required for wave equation propagation. Both 3DIFWI and 3DIFWI with M-SSIM require all spatial coordinates of the 3D Overthrust model as input, and the GPU memory required for INR forward and backward propagation is as high as 98514.4 MB. In contrast, CP- and TT-3DIFWI with M-SSIM, based on tensor decomposition strategies, require less than 20 MB of GPU memory. This result clearly demonstrates that the proposed method can significantly reduce GPU memory requirements.

\begin{table}[t]
\caption{GPU memory required for INR training.}
\centering
\begin{tabular}{lc}
\hline
Method & GPU memory \\
\hline
3DIFWI & 98514.4 MB \\
3DIFWI with M-SSIM & 98514.4 MB \\
CP-3DIFWI with M-SSIM & 7.3 MB \\
TT-3DIFWI with M-SSIM & 16.5 MB \\
\hline
\end{tabular}
\label{tab:gpu_memory}
\end{table}

Subsequently, the constant gradient velocity model is replaced by a homogeneous model as the initial model for inversion, as shown in \ref{./overthrust/result_homo.jpg}a. The number, locations, and frequency band of the seismic sources remain unchanged. The corresponding inversion results obtained using 3DFWI, 3DIFWI, 3DIFWI with M-SSIM, CP-3DIFWI with M-SSIM, and TT-3DIFWI with M-SSIM are shown in Figures \ref{./overthrust/result_homo.jpg}(b)-\ref{./overthrust/result_homo.jpg}(f), respectively. When a homogeneous model is used as the initial model, conventional 3DFWI fails to recover the correct subsurface velocity structures. As shown in Figure \ref{./overthrust/result_homo.jpg}(b), the inversion result is dominated by strong oscillatory artifacts caused by severe cycle skipping, indicating that the inversion becomes trapped in local minima due to the large discrepancy between the initial and true models. Compared with 3DFWI, 3DIFWI alleviates these artifacts to some extent by employing INR to reparameterize the velocity model. As shown in Figure \ref{./overthrust/result_homo.jpg}(c), the inversion result exhibits improved smoothness and partially recovered structural features. However, the reconstructed velocity model still contains significant structural ambiguity, and the main geological structures are not clearly resolved. By introducing the M-SSIM objective function, the inversion result of 3DIFWI is further improved. As shown in Figure \ref{./overthrust/result_homo.jpg}(d), the structural continuity becomes clearer and the major stratigraphic trends begin to emerge. Nevertheless, the complex structural region in the central part of the model remains poorly reconstructed. When tensor decomposition is incorporated into the M-SSIM-based 3DIFWI framework, the inversion quality is further improved. The CP-3DIFWI with M-SSIM result shown in Figure \ref{./overthrust/result_homo.jpg}(e) successfully reconstructs the overall layered velocity structure and significantly suppresses inversion artifacts. However, due to the limited representation capability of CP decomposition, the recovery of complex structures and detailed velocity variations remains insufficient. In contrast, TT-3DIFWI with M-SSIM (Figure \ref{./overthrust/result_homo.jpg}(f)) provides a more accurate reconstruction of both large-scale stratigraphic structures and complex structural features in the central region. Benefiting from the stronger representation capability of tensor train decomposition, this method achieves better structural continuity and more accurate velocity recovery. Overall, TT-3DIFWI with M-SSIM produces the most reliable inversion result under the homogeneous initial model condition and successfully recovers the main geological structures of the 3D Overthrust model.

To further investigate the structural reconstruction capability of different methods, orthogonal slices along the crossline, inline, and depth directions are compared in Figure \ref{./overthrust/section_result.jpg}. These slices provide a clearer view of the structural continuity and velocity variation in different spatial directions. For conventional 3DFWI, the slice results exhibit strong incoherent velocity patterns and severe structural distortions. In particular, the inline and crossline sections show highly fragmented velocity distributions, while the horizontal slice displays random velocity fluctuations. These artifacts indicate that the inversion is strongly affected by cycle skipping when the homogeneous model is used as the initial model, resulting in unstable structural reconstruction. Compared with 3DFWI, 3DIFWI produces smoother velocity distributions and partially recovers the shallow structural features. However, the deeper structures remain poorly resolved, and the complex structural region is still significantly blurred in all three slice directions. After incorporating the M-SSIM objective function, the structural continuity in both the crossline and inline sections is further improved. Some major velocity interfaces become more recognizable, and the horizontal slice shows more coherent structural patterns. Nevertheless, the recovery of deep structures and detailed velocity variations remains limited. When tensor decomposition is introduced, both CP-3DIFWI and TT-3DIFWI with M-SSIM produce significantly clearer slice images. In particular, the layered velocity structures become more continuous in the vertical sections, and the horizontal slice reveals more stable structural trends. Although CP-3DIFWI can effectively reconstruct the overall stratified velocity distribution, its ability to represent complex structural variations is still limited. In contrast, TT-3DIFWI with M-SSIM provides more accurate reconstruction of both the layered structures and the complex deformation region. The slices show improved structural continuity and clearer velocity contrasts across different layers, indicating that the tensor train decomposition offers stronger representation capability for complex subsurface structures. These results demonstrate that the proposed TT-based approach achieves more reliable structural reconstruction under the challenging homogeneous initial model condition.

\begin{figure}
\centering
\includegraphics[width=2in]{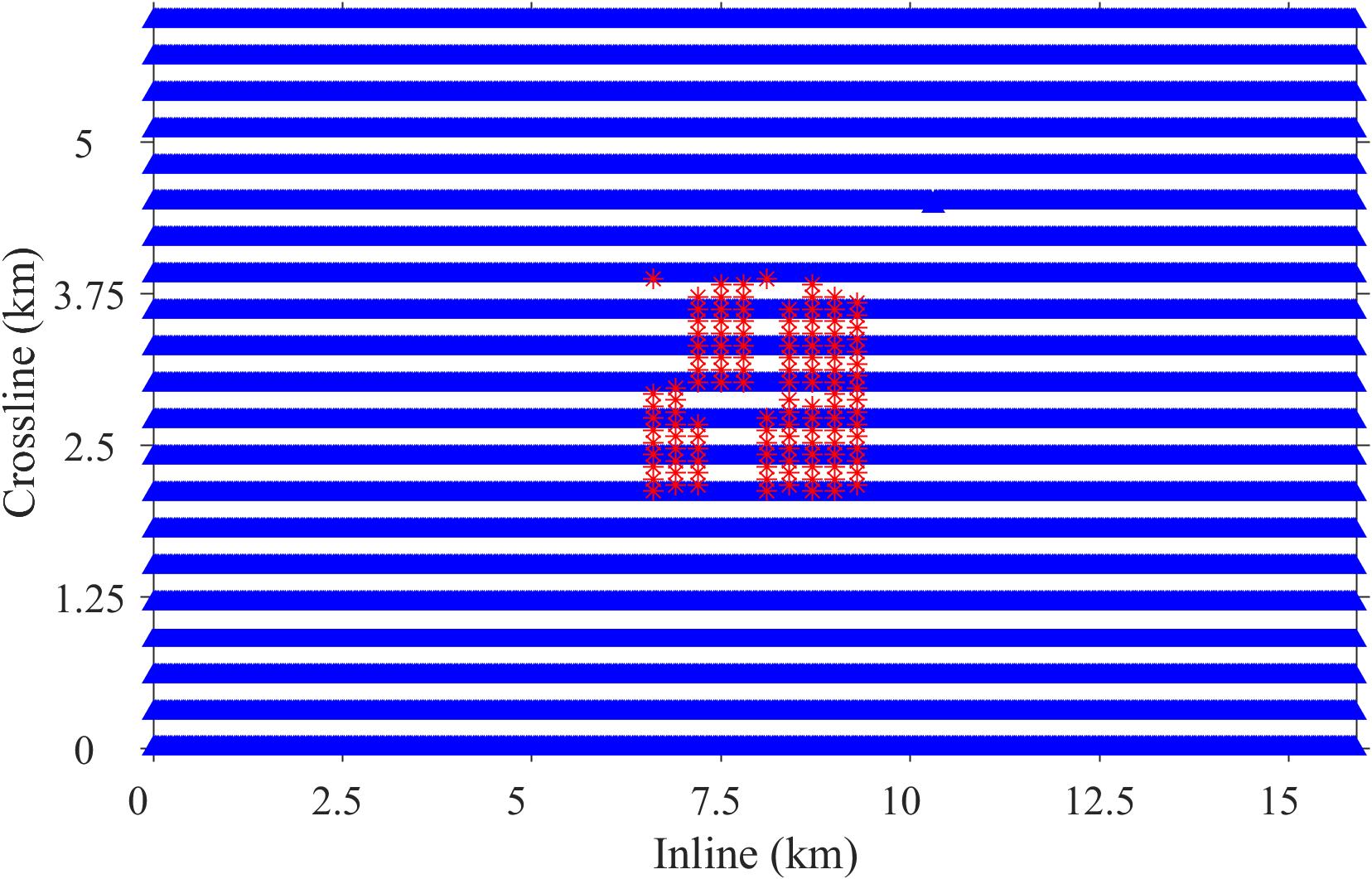}
\caption{The field data acquisition geometry.}
\label{./Field/sr_field.jpg}
\end{figure}

\begin{figure}
\centering
\includegraphics[width=3.5in]{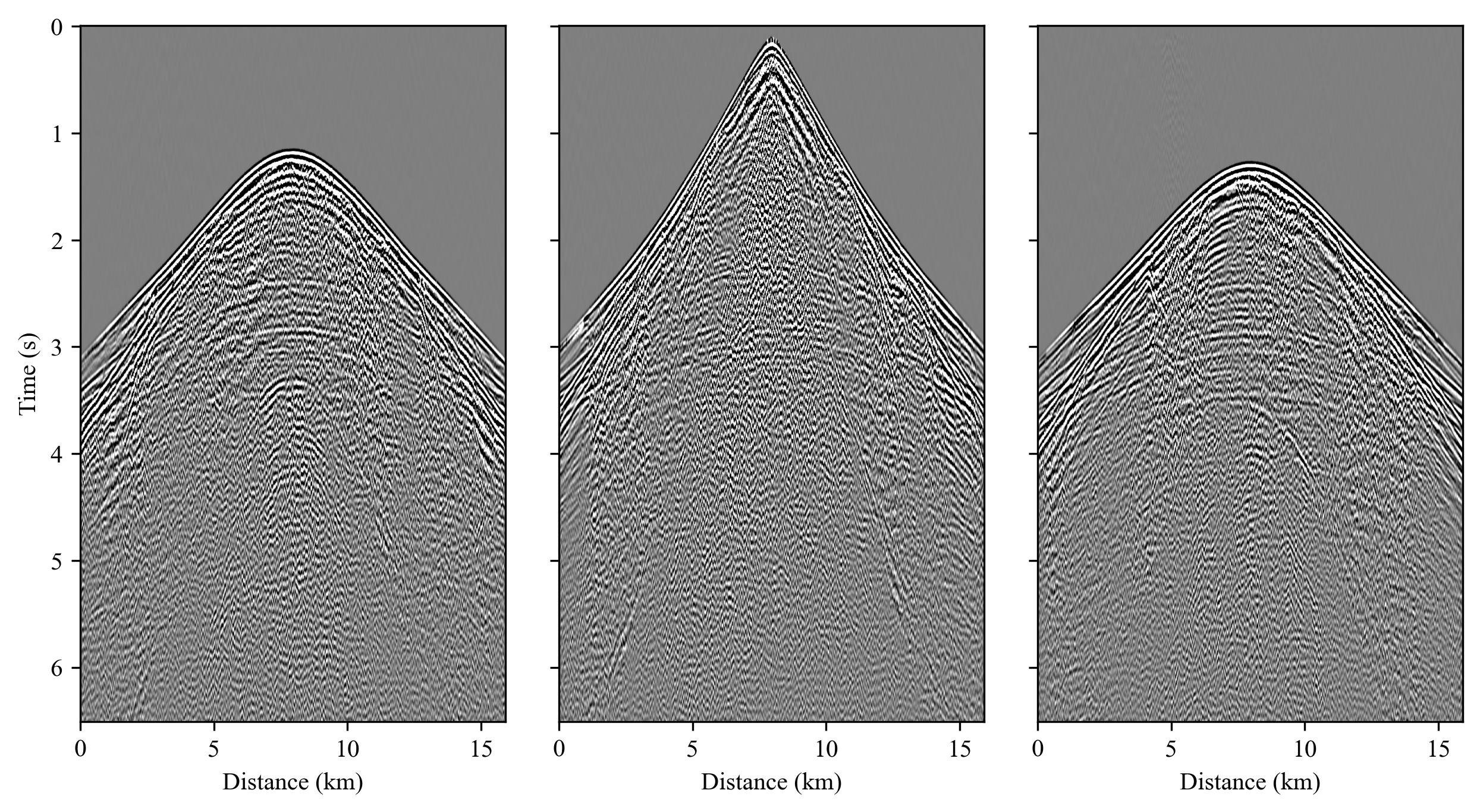}
\caption{Field data from a single shot at different Crosslines.}
\label{./Field/obs.jpg}
\end{figure}

To further provide a quantitative assessment of the velocity recovery accuracy at specific spatial positions, Figure \ref{./overthrust/Single_homo.jpg} displays the velocity–depth inversion curves at six randomly selected locations, comparing the results of different methods against the true model. For 3DFWI, the recovered velocity curves (cyan dotted lines) exhibit severe deviations from the true model (black solid lines) at almost all depths. This confirms that under a homogeneous initial model, 3DFWI is heavily trapped in local minima caused by severe cycle skipping. In contrast, 3DIFWI (purple dotted lines) yields significantly smoother velocity profiles, capturing the long-wavelength velocity variations to some extent. However, substantial discrepancies remain in regions with sharp velocity contrasts, and the reconstruction accuracy degrades rapidly with increasing depth. The incorporation of the M-SSIM objective function (green dashed lines) further aligns the inverted curves with the true model in the shallow and intermediate depth ranges. The overall velocity gradient is better preserved, although noticeable deviations persist in the deep part of the model. By integrating tensor decomposition, both CP-3DIFWI and TT-3DIFWI with M-SSIM (orange and red dashed lines, respectively) demonstrate superior performance in tracking the true velocity variations across the entire depth range. These methods effectively reconstruct the sharp velocity interfaces and maintain a consistent trend with the reference model. Notably, TT-3DIFWI with M-SSIM achieves the highest fidelity in velocity recovery among all tested methods. As observed in the profiles at (Inline=5.7 \text{ km}, Crossline=7.7 \text{ km}) and (Inline=3.0 \text{ km}, Crossline=4.5 \text{ km}), the TT-based method accurately captures the complex velocity fluctuations and shows the smallest cumulative error relative to the true model. Even in the deepest regions where other methods tend to diverge, TT-3DIFWI maintains high structural stability. These quantitative results further validate that the proposed TT-based framework, leveraging the strong representation capability of tensor train decomposition, provides the most reliable and accurate velocity reconstruction under challenging initial conditions.

\subsection{Field Data} \hfill

\begin{figure}
\centering
\includegraphics[width=1.75in]{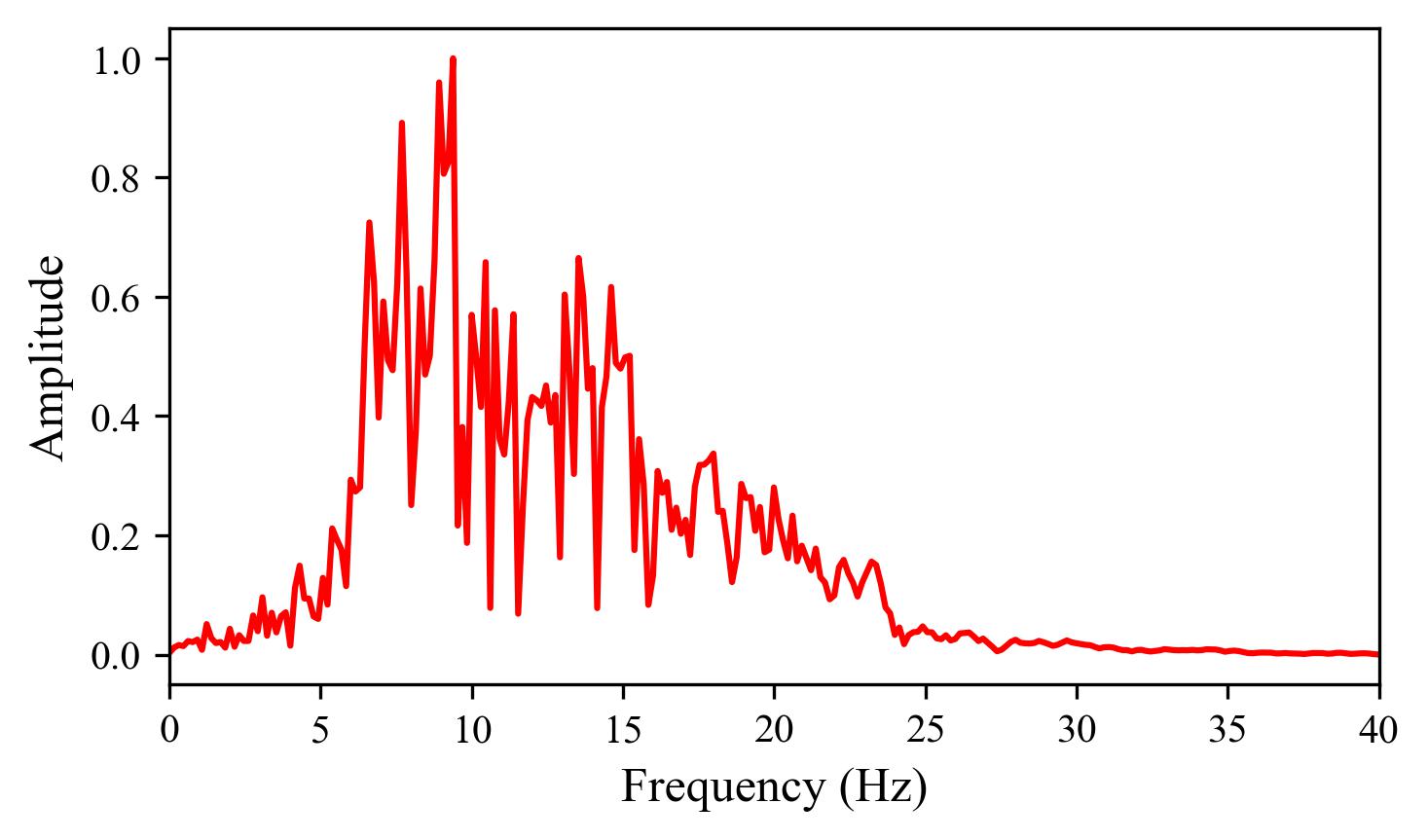}
\caption{Field data frequency spectrum.}
\label{./Field/field_f.jpg}
\end{figure}

To further validate the practical applicability of the proposed TT-3DIFWI with M-SSIM, we apply the method to a land field seismic dataset. The subsurface model is discretized into a $192 \times 242 \times 636$ grid with a uniform spacing of 25 m, covering a physical domain of approximately $ 4.8 \text{ km} \times 6 \text{ km} \times 16 \text{ km} $. A total of 124 shot gathers are extracted from the original dataset for the inversion. As illustrated in the acquisition geometry in Figure \ref{./Field/sr_field.jpg}, where red stars denote sources and blue triangles indicate receivers, each shot is recorded by an extensive array of 8,448 receivers. These receivers are arranged in a grid consisting of 16 crosslines with 528 receivers per line. The receiver spacing is 25 m in the inline direction, while the crossline spacing is considerably coarser at 300 m, resulting in a maximum offset of approximately 6.6 km. Regarding the vertical configuration, the source depths range from 20 m to 60 m, while the receiver depths vary between 10 m and 30 m. Notably, the majority of sources are concentrated in the central region of the model. This distribution leads to strong non-uniform illumination, particularly resulting in poor illumination near the model boundaries, which significantly increases the ill-posedness of the inversion. The raw seismic records consist of 1,626 time samples with a temporal sampling interval of 0.004 s, providing a total recording duration of approximately 6.5 s. These records are heavily contaminated by high-amplitude surface waves and environmental noise, which severely obscure the reflection signals. Consequently, surface-wave noise removal is applied as a mandatory preprocessing step. The processed seismic data used for the inversion are displayed in Figure \ref{./Field/obs.jpg}. Despite this preprocessing, the data still exhibit a low signal-to-noise ratio, with persistent noise components and relatively weak, fragmented reflection events. In addition, low-frequency components below 5 Hz in the data are not available, and the data frequency spectrum is shown in Figure \ref{./Field/field_f.jpg}. This combination of sparse acquisition footprints, limited boundary illumination, and complex noise interference makes the inversion task particularly challenging, providing a rigorous test for the robustness of the proposed framework.

\begin{figure}
\centering
\includegraphics[width=3.5in]{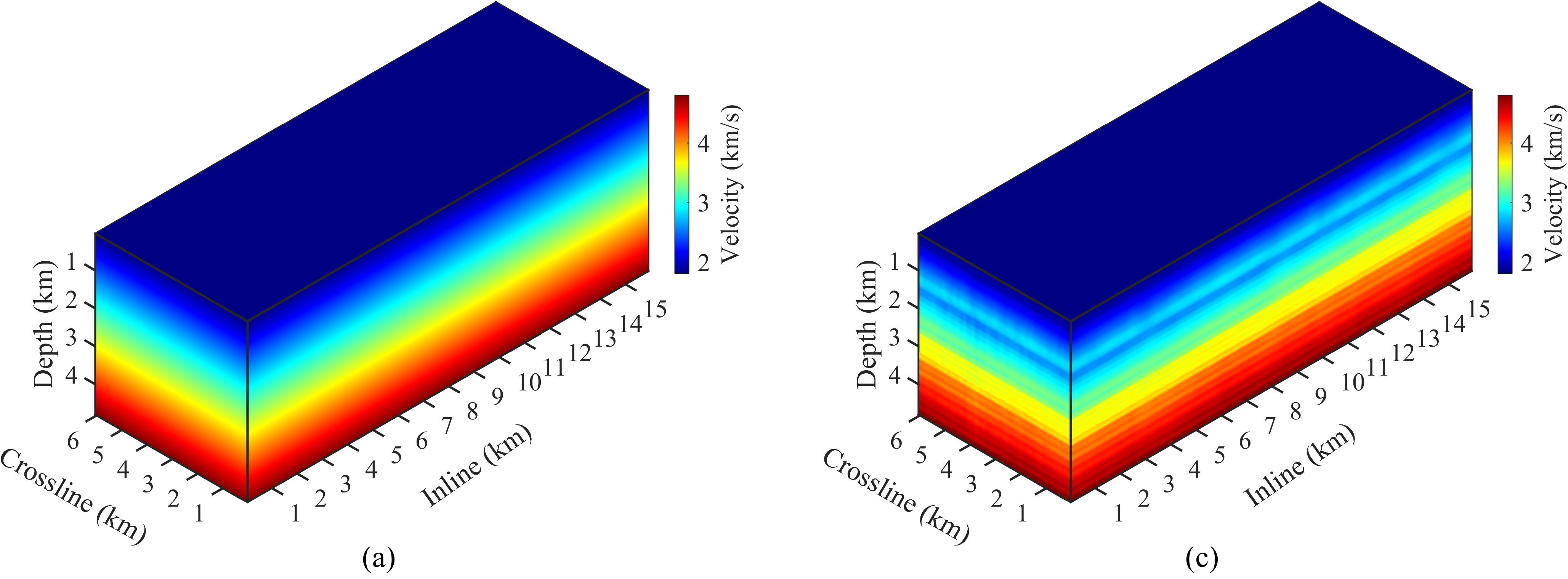}
\caption{Field data velocity model. (a) Initial model. (b) TT-3DIFWI with M-SSIM
inversion result.}
\label{./Field/3d_field.jpg}
\end{figure}

\begin{figure}
\centering
\includegraphics[width=3.5in]{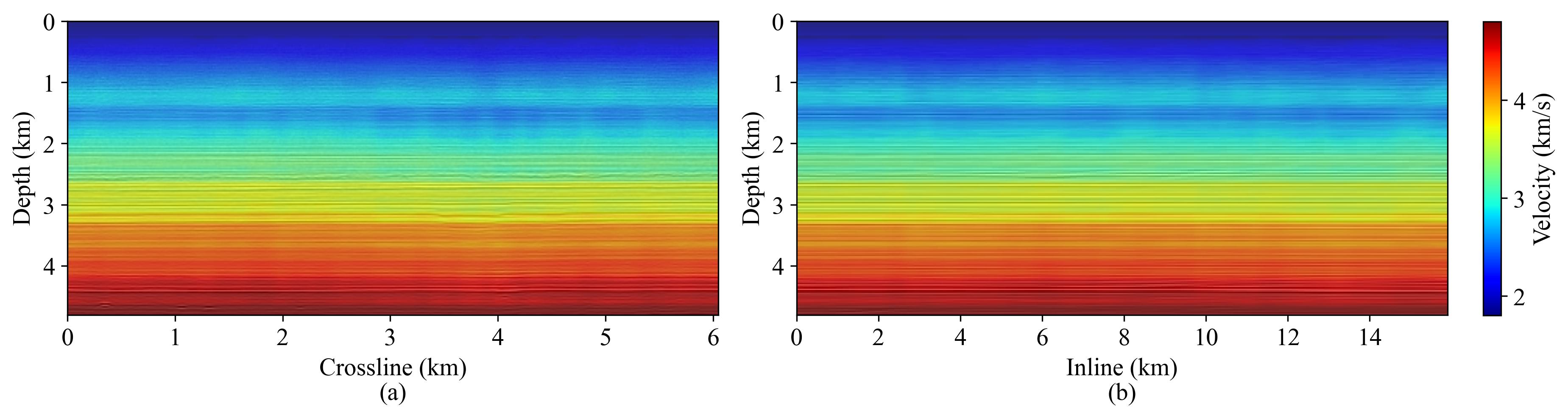}
\caption{TT-3DIFWI with M-SSIM slices inversion results from field data. (a)
Inline direction. (b) Crossline direction.}
\label{./Field/ic_field.jpg}
\end{figure}

\begin{figure}
\centering
\includegraphics[width=2.7in]{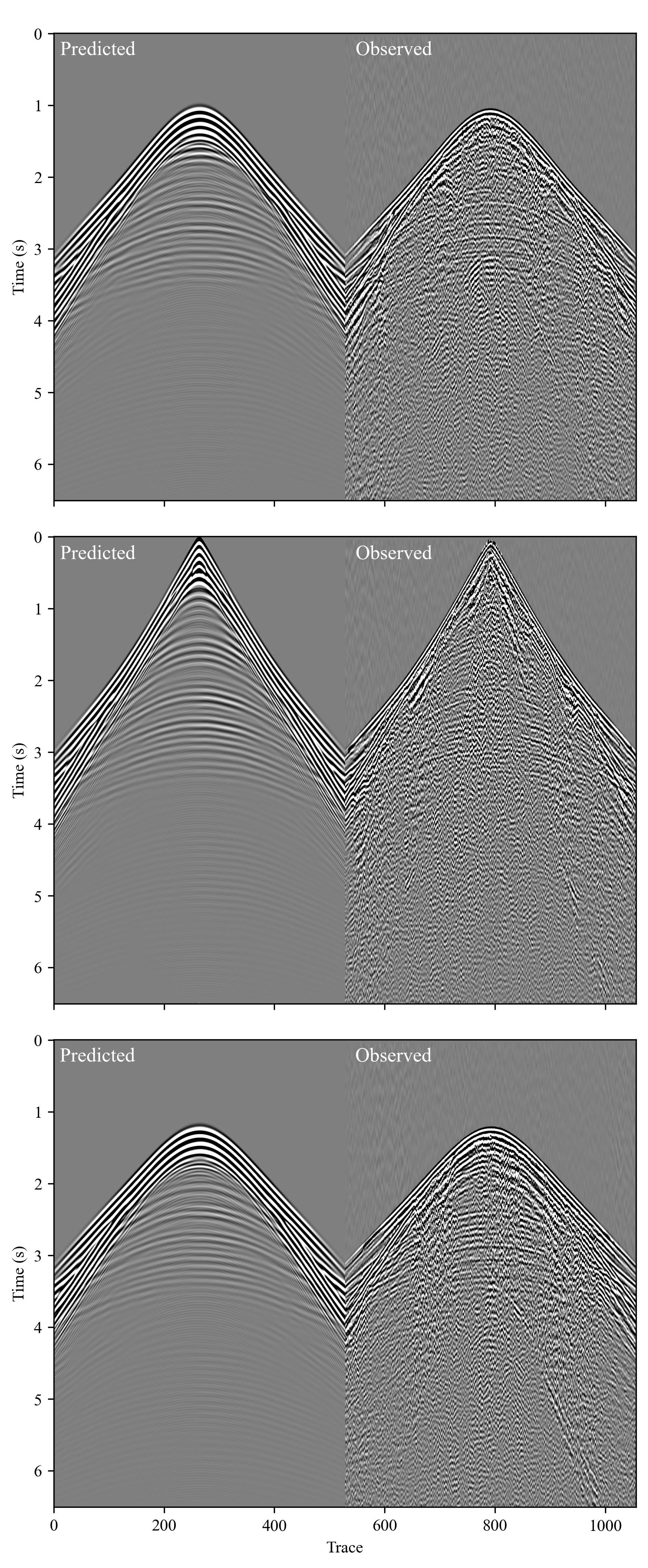}
\caption{Comparison of observed and predicted data.}
\label{./Field/pred.jpg}
\end{figure}

The initial model we are using is a constant gradient velocity model, as shown in Figure \ref{./Field/3d_field.jpg}a. The inversion result obtained by the TT-3DIFWI with M-SSIM method is shown in Figure \ref{./Field/3d_field.jpg}b, which exhibits a layered structure along the Inline direction. Subsequently, we present cross-sectional views along the crossline and inline directions, overlaid with the least-squares reverse time migration (LSRTM) result, as shown in Figure \ref{./Field/ic_field.jpg}. It is clear that the velocity inversion result and the LSRTM result have a high degree of structural consistency, both indicating a layered subsurface structure. We can see that the velocity inversion results have layered characteristics. 

The data-fitting performance is evaluated in Figure \ref{./Field/pred.jpg}, which compares the observed data with the predicted data generated from the final inverted model. We can see that the predictions match the observation data well, especially the reflections and refractions at long offsets. This high level of data consistency indicates that the inverted model effectively honors the wavefield physics recorded in the field. Finally, a quantitative validation is performed by comparing the inversion result with the available well-log data, as shown in Figure 19. The initial model (brown line) significantly deviates from the true subsurface velocity trend. In contrast, the inverted velocity profile (red line) closely tracks the low-frequency trend and major velocity variations of the well-log curve (blue line) across the entire depth range. Although the inversion cannot capture the high-frequency fluctuations inherent in log data due to the limited frequency band of seismic waves, the agreement in the background velocity and depth-dependent trends further confirms the accuracy and practical reliability of the proposed inversion framework in complex land environments.

\begin{figure}
\centering
\includegraphics[width=2.5in]{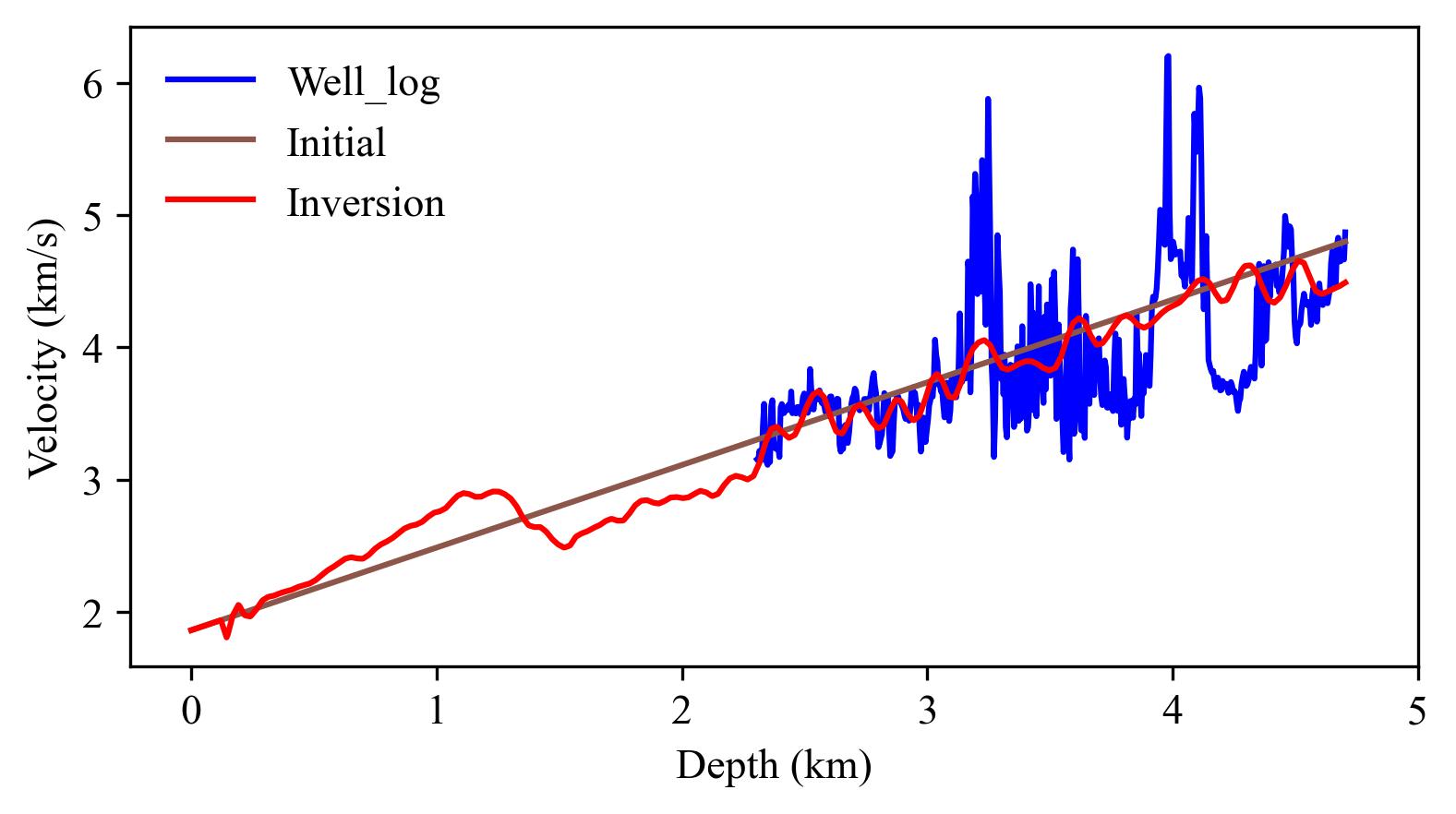}
\caption{Comparison of the well-log.}
\label{./Field/well_log.jpg}
\end{figure}

\section{Discussion} \hfill

The experimental results demonstrate that the proposed TT-3DIFWI framework can effectively reconstruct high-resolution velocity models while significantly reducing memory consumption. By representing the 3D velocity model through TT decomposition, the model parameters are factorized into a series of low-rank core tensors. This compact representation substantially decreases the number of parameters required to describe the subsurface model, which alleviates the memory burden during both forward and backward propagation of INR. Compared with conventional INR-based 3D FWI methods that directly predict the entire velocity model, the TT-based parameterization provides a more efficient representation while maintaining the reconstruction accuracy.

Another important advantage of the proposed framework lies in the structural regularization introduced by the low-rank property of the TT decomposition. The inherent low-rank constraint encourages the reconstructed velocity model to maintain consistent large-scale structural patterns, which helps suppress spurious high-frequency artifacts during inversion. As observed in the numerical experiments, the reconstructed velocity models exhibit improved structural continuity and clearer geological boundaries compared with those obtained using conventional approaches. This indicates that TT decomposition not only reduces the computational burden but also acts as an implicit regularization mechanism that stabilizes the inversion process.

In addition, the incorporation of the M-SSIM objective function further enhances the robustness of the inversion. Unlike the conventional $l_2$ norm, which relies solely on point-wise amplitude differences, M-SSIM measures structural similarity between seismic data at multiple spatial and temporal scales. This multi-scale comparison allows the inversion to focus on structural features rather than strict amplitude matching, thereby mitigating the cycle skipping problem when low-frequency information is missing or the initial velocity model is inaccurate. The misfit analysis and inversion results confirm that M-SSIM provides a smoother optimization landscape and improves convergence toward geologically meaningful solutions.

Despite these advantages, several limitations remain. The performance of TT decomposition depends on the selected TT ranks, which control the trade-off between model representation capability and memory efficiency. If the TT ranks are too low, important structural details may not be adequately represented, whereas excessively large ranks will increase the computational cost. Therefore, selecting appropriate TT ranks remains an important practical consideration. In addition, although the proposed framework significantly reduces memory requirements, the computational cost of 3D wavefield simulation remains substantial for large-scale problems. Future work will focus on further improving the efficiency and scalability of the proposed framework. Potential directions include adaptive TT-rank selection strategies, integration with more advanced wavefield compression techniques, and extension to elastic or anisotropic full waveform inversion. These developments may further enhance the applicability of TT-3DIFWI for large-scale and complex seismic imaging problems.

\section{Conclusion} \hfill

We proposed a novel tensor train-based 3D implicit full waveform inversion framework (TT-3DIFWI) combined with a multi-scale structural similarity (M-SSIM) objective function. By decomposing the 3D velocity model into low-rank core tensors, TT-3DIFWI significantly reduces memory requirements while maintaining high reconstruction accuracy. The inherent low-rank structure of the TT decomposition ensures structural consistency in the inverted velocity models, effectively preserving geological features and improving the continuity of the results. The M-SSIM objective function further enhances sensitivity to multi-scale structural differences between observed and predicted seismic data, increasing robustness against cycle skipping and missing low-frequency components. Experiments on synthetic and field datasets demonstrate that TT-3DIFWI with M-SSIM can accurately reconstruct a high-resolution velocity model under challenging conditions, including poorly initialized models and missing low-frequency data. The inversion results exhibit improved structural continuity and clearer geological features compared with conventional methods. These findings indicate that TT-3DIFWI with M-SSIM offers an effective, robust, and scalable solution for large-scale 3D seismic imaging.

\bibliography{FWI.bib}
\bibliographystyle{unsrt} 

% \begin{thebibliography}{1}

% \bibitem{kour2014real}
% George Kour and Raid Saabne.
% \newblock Real-time segmentation of on-line handwritten arabic script.
% \newblock In {\em Frontiers in Handwriting Recognition (ICFHR), 2014 14th
%   International Conference on}, pages 417--422. IEEE, 2014.

% \bibitem{kour2014fast}
% George Kour and Raid Saabne.
% \newblock Fast classification of handwritten on-line arabic characters.
% \newblock In {\em Soft Computing and Pattern Recognition (SoCPaR), 2014 6th
%   International Conference of}, pages 312--318. IEEE, 2014.

% \bibitem{hadash2018estimate}
% Guy Hadash, Einat Kermany, Boaz Carmeli, Ofer Lavi, George Kour, and Alon
%   Jacovi.
% \newblock Estimate and replace: A novel approach to integrating deep neural
%   networks with existing applications.
% \newblock {\em arXiv preprint arXiv:1804.09028}, 2018.

% \end{thebibliography}

\end{document}